\documentclass[sigplan,twocolumn]{acmart}

\usepackage{xspace}
\usepackage{subfigure}
\usepackage{url}
\usepackage{zref-totpages}

\usepackage{wrapfig}

\usepackage{hyperref}
\makeatletter
\def\UrlAlphabet{%
      \do\a\do\b\do\c\do\d\do\e\do\f\do\g\do\h\do\i\do\j%
      \do\k\do\l\do\m\do\n\do\o\do\p\do\q\do\r\do\s\do\t%
      \do\u\do\v\do\w\do\x\do\y\do\z\do\A\do\B\do\C\do\D%
      \do\E\do\F\do\G\do\H\do\I\do\J\do\K\do\L\do\M\do\N%
      \do\O\do\P\do\Q\do\R\do\S\do\T\do\U\do\V\do\W\do\X%
      \do\Y\do\Z}
\def\UrlDigits{\do\1\do\2\do\3\do\4\do\5\do\6\do\7\do\8\do\9\do\0}
\g@addto@macro{\UrlBreaks}{\UrlOrds}
\g@addto@macro{\UrlBreaks}{\UrlAlphabet}
\g@addto@macro{\UrlBreaks}{\UrlDigits}
\makeatother

\usepackage{tikz}

\usepackage{amsmath}

\newcommand{\name}{\textsc{Palladium}\xspace}

\newcommand{\skmsg}{\texttt{SK\_MSG}\xspace}
\newcommand{\hide}[1] {}

\newcommand{\ie}{{\em i.e., \/}}
\newcommand{\eg}{{\em e.g., \/}}

\definecolor{lincolngreen}{rgb}{0.11, 0.35, 0.02}
\definecolor{majorelleblue}{rgb}{0.38, 0.31, 0.86}

\newcommand{\cmark}{\color{blue} \ding{51}}%
\newcommand{\xmark}{\color{red} \ding{55}}%

\usepackage[normalem]{ulem}
\usepackage{comment}

\newcommand{\textnb}[1]{\noindent\textbf{#1}}
\newcommand{\textnit}[1]{\noindent\textit{#1}}

\newcommand{\textib}[1]{\textbf{\textit{#1}}}

\usepackage{listings}

\definecolor{codegreen}{rgb}{0,0.6,0}
\definecolor{codegray}{rgb}{0.5,0.5,0.5}
\definecolor{codepurple}{rgb}{0.58,0,0.82}
\definecolor{backcolour}{rgb}{0.95,0.95,0.92}

\lstdefinestyle{mystyle}{
 backgroundcolor=\color{backcolour},
 commentstyle=\color{codegreen},
 keywordstyle=\color{magenta},
 numberstyle=\tiny\color{codegray},
 stringstyle=\color{codepurple},
 basicstyle=\ttfamily\footnotesize,
 breakatwhitespace=false,         
 breaklines=true,                 
 captionpos=b,                    
 keepspaces=true,                 
 numbers=left,                    
 numbersep=2pt,                  
 showspaces=false,                
 showstringspaces=false,
 showtabs=false,                  
 tabsize=2
}

\newcommand{\X}{$\times$\xspace}

\usepackage[font=small]{caption}

\usepackage{graphicx}
\usepackage{multirow}

\usepackage{pifont}

\usepackage{enumitem}
\setlist[enumerate]{leftmargin=0.5cm}
\setlist[itemize]{leftmargin=0.4cm}

\usepackage[T1]{fontenc}
\usepackage[breakable, theorems, skins]{tcolorbox}
\tcbset{enhanced}

\DeclareRobustCommand{\mybox}[2][gray!20]{%
\begin{tcolorbox}[   
        breakable,
        left=1pt,
        right=1pt,
        top=1pt,
        bottom=1pt,
        colback=#1,
        colframe=#1,
        width=\dimexpr\linewidth\relax, 
        enlarge left by=0mm,
        boxsep=1pt,
        arc=0pt,outer arc=0pt,
        ]
        #2
\end{tcolorbox}
}

\begin{document}


\renewcommand\footnotetextcopyrightpermission[1]{} 
\setcopyright{none}

\settopmatter{printacmref=false, printccs=false, printfolios=true}

\acmDOI{}
\acmISBN{}

\title{Palladium: A DPU-enabled Multi-Tenant Serverless Cloud over Zero-copy Multi-node RDMA Fabrics}

\author{\fontsize{11}{10}\selectfont
Shixiong Qi$^\triangle$, Songyu Zhang$^\dagger$, K. K. Ramakrishnan$^\dagger$, Diman Z. Tootaghaj$^\star$, Hardik Soni$^\star$, Puneet Sharma$^\star$\\
$^\triangle$University of Kentucky, $^\dagger$University of California, Riverside, $^\star$Hewlett Packard Labs}

\renewcommand{\shortauthors}{Qi et al.}
\renewcommand{\shorttitle}{Palladium: A DPU-enabled Multi-Tenant Serverless Cloud}
\renewcommand{\authors}{S. Qi, S. Zhang, K. K. Ramakrishnan, D. Z. Tootaghaj, H. Soni, P. Sharma}

\begin{abstract}

Serverless computing promises enhanced resource efficiency and lower user costs, yet is burdened by a heavyweight, CPU-bound data plane. Prior efforts exploiting shared memory reduce overhead locally but fall short when scaling across nodes. Furthermore, serverless environments can have unpredictable and large-scale multi-tenancy, leading to contention for shared network resources.

We present Palladium, a DPU-centric serverless data plane that reduces the CPU burden and enables efficient, zero-copy communication in multi-tenant serverless clouds.
Despite the limited general-purpose processing capability of the DPU cores, Palladium strategically exploits the DPU's potential by (1) offloading data transmission to high-performance NIC cores via RDMA, combined with intra-node shared memory to eliminate data copies across nodes, and (2) enabling cross-processor (CPU-DPU) shared memory to eliminate redundant data movement, 
which overwhelms wimpy DPU cores.
At the core of Palladium is the DPU-enabled network engine (DNE) -- a lightweight reverse proxy that isolates RDMA resources from tenant functions, orchestrates inter-node RDMA flows, and enforces fairness under contention.

To further reduce CPU involvement, Palladium performs early HTTP/TCP-to-RDMA transport conversion at the cloud ingress, bridging the protocol mismatch before client traffic enters the RDMA fabric, thus avoiding costly protocol translation along the critical path.
We show that careful selection of RDMA primitives (i.e., two-sided instead of one-sided) significantly affects the zero-copy data plane.

Our preliminary experimental results show that enabling DPU offloading in Palladium improves RPS by 20.9\X. The latency is reduced by a factor of 21\X in the best case, all the while saving up to 7 CPU cores, and only consuming two wimpy DPU cores.

\end{abstract}

\maketitle

\section{Introduction}\label{sec:intro}

Serverless computing, or Function-as-a-Service (FaaS~\cite{faas}), 
eases the burden on application developers to provision and manage cloud resources, while its fine-grained resource elasticity dramatically reduces user costs~\cite{berkeley-serverless, xfaas, infinicache}.
Despite many positive features, 
the data plane in current serverless platforms is heavily \textit{CPU-bound}, caused by significant overheads from kernel-based inter-function networking~\cite{spright, nightcore, boucher-atc18, granular-computing-hotos19}.
This cost is further amplified by the loose coupling between serverless functions, resulting in duplicate data plane processing in a ``chain'' of disaggregated functions and severely impacting performance.

To reduce the data plane overhead in serverless environments, existing solutions have leveraged shared memory processing, enabling faster and efficient {\it zero-copy} communication between functions~\cite{spright, ditto, pheromone,yuanrong}. However, they are primarily for intra-node communication, restricting their applicability in a distributed serverless system where functions of a chain may be spread across multiple nodes.
RDMA-based network fabrics, widely deployed in data centers~\cite{aquila, azure-rdma-ndsi23}, can extend zero-copy communication\footnote{
In our context, ``zero-copy'' specifically refers to the elimination of software-based data copies, while allowing hardware-based mechanisms like DMA/RDMA for efficient data movement.} to distributed environments and enable low-latency, efficient inter-function communication by leveraging hardware acceleration~\cite{farm, rmmap, fuyao}.
More recently, RDMA NICs (RNICs) have been enhanced with onboard, general-purpose processing cores, evolving into SmartNIC-like Data Processing Units (DPUs)~\cite{xingda, bluefield}. DPUs can offload and orchestrate data plane tasks directly on the integrated RNICs and have become a fundamental element of compute nodes in the cloud~\cite{BurstCBS, alibaba-cloud-cipu, baidu-cloud-snic, alibaba-luna-solar-system, AccelNet}.
Despite these infrastructure-level advancements, several challenges remain to be addressed before its full potential can be unleashed in serverless environments:

Multi-tenancy related challenges persist in the serverless environment, with resource contention 
becoming even more dynamic and unpredictable. 
This stems from the fundamental goal of fine-grained resource elasticity of serverless computing. 
Frequent configuration changes due to workload variation, function placement and auto-scaling require corresponding flexibility in provisioning of compute/network resources for each tenant.
While existing platforms have developed robust CPU, memory, and traditional network I/O isolation across tenants~\cite{faasm, sigmaos, particle}, they largely overlook RDMA-related network isolation, leaving inter-tenant RDMA traffic vulnerable to interference.
This gap exists because serverless functions typically lack the privilege to access RNIC directly (i.e., operating RDMA Queue Pairs (QPs) via IB verbs).
Granting untrusted user functions direct access to RDMA resources 
can introduce significant security risks~\cite{freeflow, masq, lite} and fail to ensure fair sharing (\eg of the bandwidth of the RDMA Fabric) among different tenants (more details in \S\ref{sec:challenges}).

Serverless platforms strive to maximize the density of user functions per node~\cite{alibaba-rund, firecracker}.
However, having the CPU, which is the primary place for function execution, perform auxiliary orchestration tasks (such as managing multi-tenancy and operating RDMA QPs) consumes valuable CPU cycles, 
which instead could be used for executing serverless functions. 
DPUs present an attractive offloading target, but face two principal limitations:
\textit{(1) Wimpy DPU cores:} DPU cores are generally much less powerful than those of CPUs, so they struggle with heavyweight packet processing~\cite{ipipe, alibaba-luna-solar-system}, especially for protocol processing and state management on the DPU. 
\textit{(2) Cross-processor (CPU-DPU) data movement overhead:} 
To perform network orchestration on the DPU, inbound packets must traverse from the RNIC to the DPU, before onward to the CPU for function execution, and vice versa for outbound packets. This ``on-path'' CPU–DPU data movement incurs nontrivial overhead, despite being accelerated by the DPU's DMA engine. Empirical measurements (see \S\ref{eval:x-processor-shm}) show up to a 1.54\X performance degradation, compared to having the data directly transmitted from host (CPU) out to the network using the RNIC's DMA.

To create a zero-copy dataplane in a distributed environment, 
inter-node (RDMA) and intra-node (shared memory) data transfers should share a \textit{unified memory pool}. Maintaining separate pools for inter-node and intra-node communication inevitably results in data copies (Fig.~\ref{fig:RDMA-primitives-selection} (2)). But sharing a unified memory pool for a combined inter-node and intra-node dataplane is nontrivial. Since the memory pool and associated data buffers are shared, careful management of data ownership is essential to prevent contention. Although locks and collective synchronization (such as Rendezvous~\cite{rendezvous-mpi}) can be used to manage data ownership between distributed functions, they can also significantly degrade performance (Fig.~\ref{fig:RDMA-primitives-selection} (1)).

Additionally, external clients use HTTP/TCP while the RDMA Fabric uses a more lightweight transport.
Existing RDMA-based serverless data planes often overlook this transport mismatch, still relying on HTTP/TCP ingress gateways at the cluster edge to forward TCP connection traffic from external clients to worker nodes.
The worker's local TCP/IP stack has to process requests before switching to RDMA or shared memory (see Fig.~\ref{fig:transport-protocol-adapation}). 
This redundant processing at both the ingress and worker nodes leads to inefficiencies. Our findings (\S\ref{eval:cluster-ingress}) show up to 11.4\X performance degradation and also significant CPU consumption.
We believe the TCP connection ought to be properly terminated at the ingress gateway, and only the payload efficiently transferred over RDMA to maximize performance.

\textnb{\name Overview.}
To address these concerns, we propose \name, a DPU-centric serverless data plane 
that seamlessly integrates intra-node shared memory processing with inter-node RDMA-based data transfers, interacting over a unified memory pool on each worker node. This facilitates {\it true} zero-copy data transfers between functions and eliminates the heavyweight kernel-based data transfers, regardless of their physical location within the serverless cloud.

To support multi-tenancy in serverless environments using RDMA Fabrics, \name introduces a node-wide, DPU-enabled Network Engine (called ``DNE'') as a software indirection layer to operate RDMA QPs on behalf of user functions. This allows fast inter-node communication while isolating the RNIC from direct access by untrusted functions. This centralized management of a node's RDMA resources enforces per-tenant traffic management policies, ensuring fair sharing of RDMA resources (\eg bandwidth) among tenants (\S\ref{sec:multi-tenancy}).
In addition, DPU offloading offers two key benefits: (1) isolating the trusted network engine \textit{physically} from untrusted user functions running on the CPU~\cite{cpuless-system, nvidia-dpu-isolation, dpu-hotchips33} to maintain its integrity, while (2) mitigating any CPU consumption for executing network engine tasks. 

\textnb{Unleashing DPU Offloading with \name.}
To unleash the power of the DPU, despite its wimpy processor cores, 
\name incorporates several key innovations into the DNE:
\textit{(1) Cross-processor memory sharing} (\S\ref{sec:x-processor-shm}).
Shared memory between the DPU and the CPU eliminates the need (and associated cost) for ``on-path'' data movement made by the DNE.
The DNE only serves as an \textit{off-path} controller for managing RDMA data transfers and ensuring fair sharing of RNIC resources in a multi-tenant environment.
\textit{(2) Run-to-completion packet processing} (\S\ref{sec:dne}).
The DNE employs a non-blocking, run-to-completion I/O model to handle two-sided RDMA operations and support multi-tenancy,
inspired by the approach taken by IX~\cite{ix-osdi} and demikernel~\cite{demikernel}. 
This minimizes context switching and CPU scheduling overheads, alleviating the performance constraints of the wimpy DPU cores.

To address both the memory management \& synchronization challenge and the mismatch between HTTP/TCP vs. RDMA transport protocol mismatch, we introduce two key innovations in \name's zero-copy data plane:
\textit{(1) Lock-free communication primitives} (\S\ref{sec:datapath-design}).
\name selects two-sided RDMA primitives to avoid the need for distributed locks or additional copies for inter-node communication, which are inevitable when using one-sided RDMA (see discussion in \S\ref{sec:challenges}). In intra-node data plane, we adopt a `token-passing' scheme for ownership transfer of shared memory buffers.
Both mitigate the need of locks and avoid the heavyweight synchronization schemes like Rendezvous.
\textit{(2) Early transport conversion at cluster ingress} (\S\ref{sec:cluster-ingress}). 
By performing HTTP/TCP-to-RDMA conversion at the cluster-wide ingress (the earliest entry point into the serverless cloud), \name removes all software-based protocol processing overheads from the worker nodes.
This aligns perfectly with \name's zero-copy design principle.
\name integrates the DPDK-based F-Stack~\cite{f-stack} with the cluster-wide ingress for high-performance TCP/IP protocol processing and uses horizontal scaling to dynamically accommodate DPDK's polling cost.

The major contributions of \name include:
\setlist{nolistsep}\begin{itemize}[noitemsep]
    \item Real-world workload evaluations (\S\ref{sec:real-workload}) show that despite DPU cores being far less powerful than CPU cores, \name fully harnesses the DPU offloading,
    achieving up to 20.9\X throughput improvement compared to different state-of-the-art serverless data planes, while having very low latency. 
    DPU offloading in \name can save up to 7 CPU cores, only consuming two wimpy DPU cores.

    \item We show that having the serverless data plane built on top of two-sided RDMA yields up to 2.8\X reduced latency and up to 2.7\X improved throughput compared to variants using one-sided RDMA (\S\ref{eval:rdma-primitives}).

    \item \name exploits the DPU in the serverless data plane to manage multi-tenancy in RDMA Fabrics. Evaluation in \S\ref{eval:multi-tenancy} shows that \name can fairly and precisely share RNIC bandwidth between tenants according to their weights.

    \item We introduce the first high-performance HTTP/TCP-to-RDMA ingress gateway designed for serverless environments. Compared to a traditional HTTP/TCP-based ingress gateway, even those using a high-performance TCP/IP stack, our design achieves a 3.4\X reduction in end-to-end latency and a 3.2\X increase in throughput.

\end{itemize}

\section{Background \& Motivation}\label{sec:background}

    \begin{figure}[t]
    \centering
        \includegraphics[width=\columnwidth]{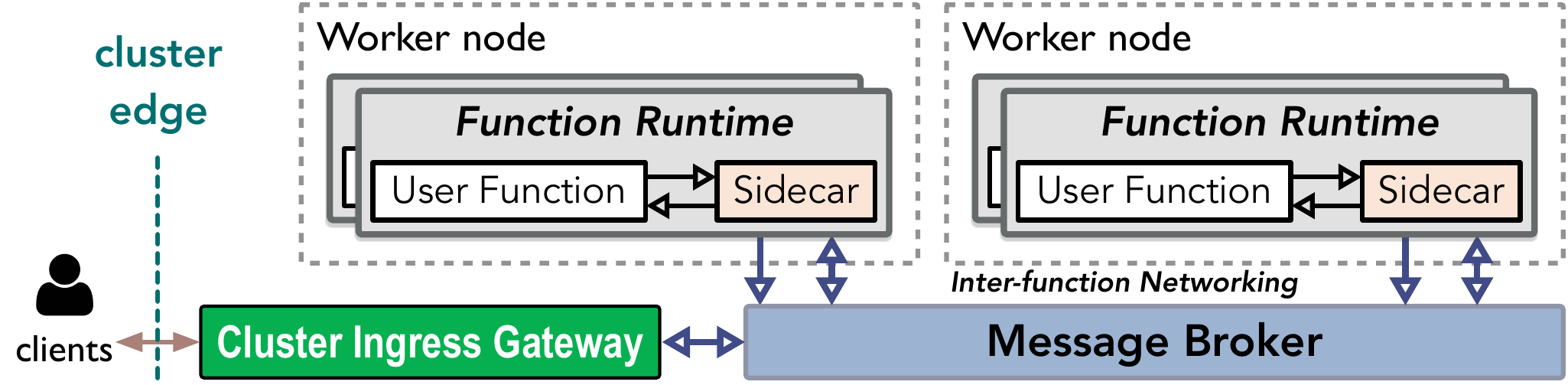}
    \vspace{-7mm}
    \caption{The architecture of serverless data plane (based on Knative's Serving/Eventing architecture~\cite{knative}).}
    \label{fig:serverless-data-plane-overview}
    \vspace{-5mm}
    \end{figure}

    \begin{figure*}[htbp]
    \centering
        \includegraphics[width=\linewidth]{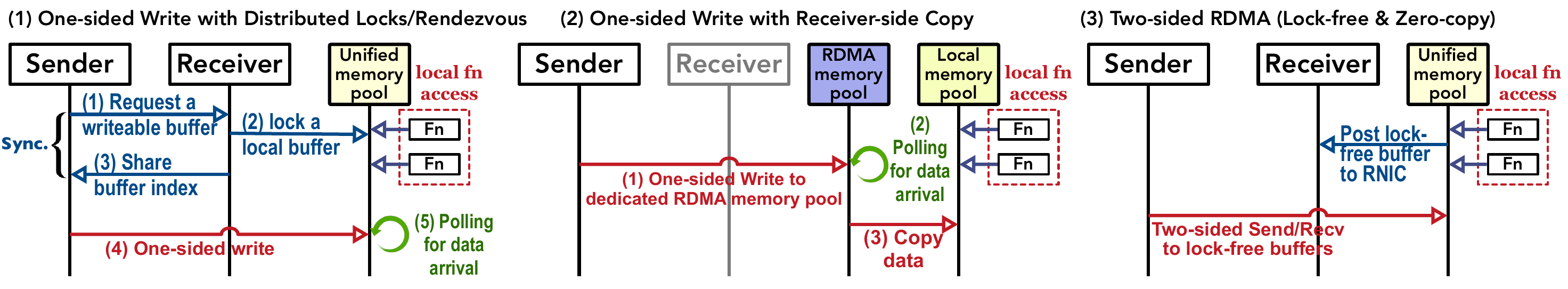}
    \vspace{-8mm}
    \caption{Selection of RDMA primitives for lock-free zero-copy serverless data plane.}
    \label{fig:RDMA-primitives-selection}
    \vspace{-4mm}
    \end{figure*}

\textnb{A Quick Primer on Serverless Data Planes.}
Fig.~\ref{fig:serverless-data-plane-overview} shows the key components of a typical serverless data plane.
Each serverless function comprises 
a {\it user function container} responsible for executing the application logic and a separate {\it sidecar container}~\cite{knative, istio} for executing 
service mesh functionality, which helps to orchestrate the loosely coupled serverless functions.
At the edge of this infrastructure is a \textit{cluster-wide ingress gateway}, which serves as the entry point to the serverless cluster, facilitating tasks such as authentication.
To facilitate invocations between serverless functions (when organized as function chains), serverless platforms generally rely on a \textit{message broker} or \textit{coordinator}.
These data plane components typically communicate using {\it slow, kernel-based networking}, which incurs significant overheads, including data copies, context switches, interrupts, protocol processing (e.g., TCP/IP) and serialization/deserialization~\cite{qizhe-21}.

\textnb{RDMA enables zero-copy networking to be distributed.}
RDMA facilitates {\it zero-copy}, {\it kernel-bypass} data movement between nodes equipped with RDMA-capable NICs. This capability extends
zero-copy communication
across multiple nodes in a serverless environment. It allows the widely used intra-node shared memory approaches to scale, without having to rely on the limitations of locality-aware placement strategies to maximize performance~\cite{ditto, spright, pheromone, wukong}. Such placement strategies are often impractical due to node-wide resource constraints and the massive scale of production cloud applications~\cite{alibaba-trace-analysis}.
Moreover, while a zero-copy userspace TCP/IP stack~\cite{sure, demikernel, mrpc} also enables \textit{zero-copy}, \textit{kernel-bypass} communication across nodes, RDMA still stands out, benefited by the offloading of transport-layer processing from the CPU to hardware.
This is further enabled by the widespread adoption of RDMA in cloud infrastructures.

\subsection{Challenges and Design Decisions}\label{sec:challenges}

We identify RDMA as the ideal choice to enable a distributed zero-copy data plane 
for serverless computing.
We focus on Reliable Connected (RC) RDMA transport in this work, ensuring in-order delivery and end-to-end reliability for application-layer transactions. This is also commonly used in practice~\cite{azure-rdma-ndsi23, farm}.
Despite early explorations~\cite{rmmap, fuyao}, several challenges remain to be addressed.

\textnb{Challenge\#1. Serverless computing lacks multi-tenancy support for RDMA Fabrics.}
Applications interact with the RNIC for data transfer via the QP abstraction~\cite{rdma-design-guidelines}, which requires privileged, low-level control of the RNIC.
However, existing serverless platforms typically block low-level privileges from user code due to security concerns.
Further, direct access of QPs by serverless functions is also undesirable since it lacks the required isolation of shared RDMA resources across different tenants.
Functions from different tenants would have to contend for locks to access the shared QP resources~\cite{flor-osdi, flock-sosp21, farm, rdma-design-guidelines}.
This impacts performance due to lock contention as well as frequent cache line thrashing for QP buffers across CPU cores~\cite{rdma-design-guidelines, farm}.
Further, it is more difficult to enforce a fair allocation of QPs to tenants or prioritize traffic of important tenants. E.g., a `rogue' tenant could occupy a set of QPs for a long time (i.e., possibly even deliberately not release the lock), thus starving other tenants of RDMA resources.
Thus, there is no safe way for untrusted user code to exploit RDMA's high-performance data paths. Instead, RDMA must be offered as a managed network service, fully isolated from user-level code.

\vspace{-2mm}
\mybox{\textit{Design Implication\#1}:
Exposing RDMA QPs directly to user functions is less feasible (and harmful) in a multi-tenant serverless environment. A managed network service (such as a node-wide network engine) helps enforce tenant-based isolation and manage QPs on behalf of user functions.}
\vspace{-2mm}

    \begin{figure}[b]
    \vspace{-7mm}
    \centering
        \includegraphics[width=\linewidth]{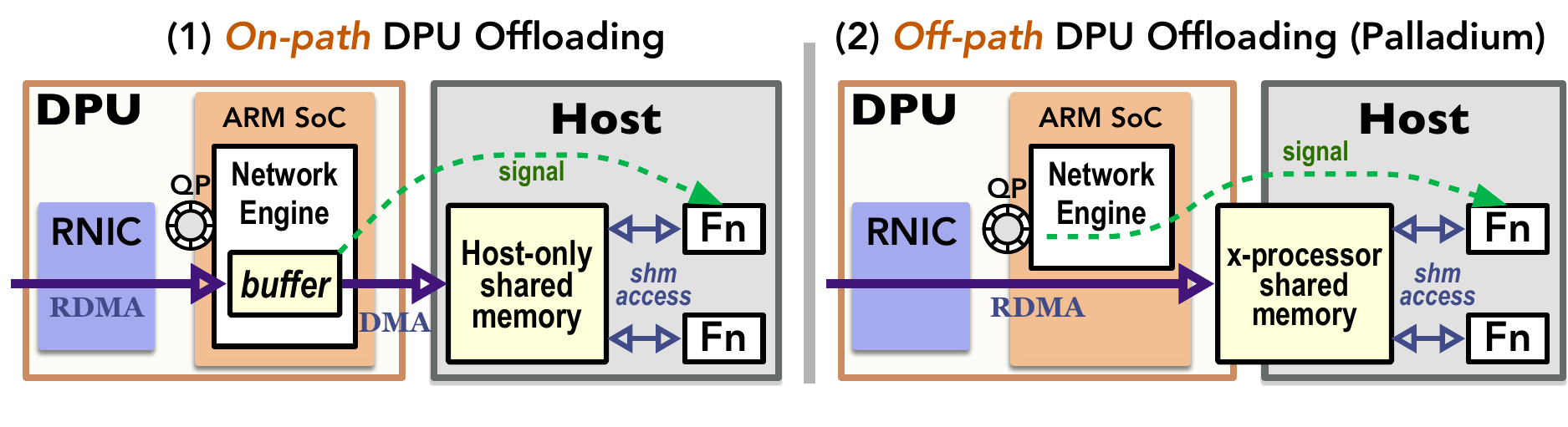}
    \vspace{-8mm}
    \caption{(1) On-path DPU offloading VS. (2) off-path DPU offloading (using cross-processor shared memory).}
    \label{fig:x-processor-shm}
    \end{figure}

\textnb{Challenge\#2: Wimpy DPU Cores Slow Down On-path Processing when DPU Offloading.}
The DPU SoC can operate in the \textit{on-path} mode or \textit{off-path} mode~\cite{ipipe, xingda}. 
\textit{On-path} mode allows full access to the in-flight data (since data is buffered in the DPU before being transferred further, as in Fig.~\ref{fig:x-processor-shm} (\textbf{1})). 
This enables the network engine on the DPU SoC to perform lightweight
network processing, \eg for multi-tenancy support, before the data is relayed to the host functions.
However, the on-path mode incurs additional data movement 
between the network engine on the DPU and the host. Even using the SoC DMA (which we find to be unfortunately very slow) can slow down the data path by up to \textbf{1.33\X} as we assess in \S\ref{eval:x-processor-shm}.

DPU SoCs configured in \textit{off-path} mode are typically considered to have limited (or no) access to the packet data, as the data is moved directly between the integrated RNIC and host memory~\cite{ipipe}. Thus, under the \textit{off-path} mode, it is very difficult to have the equivalent network protocol processing on the DPU as the \textit{on-path} mode.
However, by creating a unified, cross-processor shared memory between the DPU and the host (CPU), we can expose the buffers of the host to the DPU \textit{even in the off-path} mode (see Fig.~\ref{fig:x-processor-shm} (\textbf{2})).
This is made possible by having a cross-processor memory map primitive on the DPU (e.g., DOCA mmap library for the NVIDIA Bluefield DPU~\cite{doca}). 
The integrated RNIC on the DPU performs RDMA operations to {\it directly} transfer data into the host shared memory.
The \textit{off-path} network engine running on the DPU manages the inter-node RDMA data paths by manipulating the RDMA QPs, without adding delays to the data path.
The data is moved by the RNIC DMA (which runs at line rate) directly into the host memory, without encountering the limitation of the slow SoC DMA.

\vspace{-2mm}
\mybox{\textit{Design Implication\#2}: 
Cross-processor shared memory enhances \textit{off-path} DPU offloading with full control of the data buffer while eliminating the additional data movement between DPU and the host (CPU) of the \textit{on-path} mode.}

\textnb{Challenge\#3: Selection of RDMA primitives for lock-free zero-copy communication.}
RDMA supports two modes of operation: one-sided and two-sided~\cite{rdma-design-guidelines}. One-sided operations allow direct read or write access to buffers in the pre-registered memory region on a remote node without requiring the active involvement of the remote CPU. 
In contrast, two-sided operations adhere to traditional message-passing semantics, where both the sender and receiver actively participate in the communication. The sender delivers the data to a buffer explicitly posted by the receiver, offering greater flexibility in handling dynamic memory access requirements.
There is no one-size-fits-all answer to whether one-sided or two-sided RDMA operation is better~\cite{prism-sosp21}.
Conventional wisdom is that one-sided operations
are faster than two-sided operations, while being highly CPU efficient~\cite{farm}.

The core of a lock-free zero-copy data plane in a distributed serverless environment lies in having a {\it unified memory pool} on each node.
Because functions may execute asynchronously across nodes, simultaneous inter-node RDMA access and intra-node shared-memory accesses to a buffer in the {\it unified memory pool} can conflict, resulting in data races.

However, one-sided RDMA operations have a fatal flaw for our application scenario, 
which we define as the sender being ``receiver-oblivious'': The sender is not aware of whether the receiver's buffer is currently accessible for remote read or write, {\it since a function on the receiver node may already be using it},
also leading to potential data races.

Two potential workarounds exist: (1) employ distributed locks~\cite{citron, alock, dslr-sigmod18} or Rendezvous-based synchronization~\cite{rendezvous-mpi} to coordinate remote one-sided RDMA read/write operations with local shared memory processing (Fig.~\ref{fig:RDMA-primitives-selection} (1)). However, this incurs additional synchronization overhead across nodes. or (2) 
dedicating an RDMA-only memory pool on the receiver node, exclusively used for remote one-sided operations (Fig.~\ref{fig:RDMA-primitives-selection} (2)), which avoids data races by isolating it from the memory pool shared by local functions. However, this introduces an additional data copy, which undermines the goal of having an efficient zero-copy design.

\textnb{Why two-sided RDMA is a better fit.}
Two-sided RDMA operations inherently avoid the sender being ``receiver-oblivious'' (see Fig.~\ref{fig:RDMA-primitives-selection} (3)). The receiver explicitly posts a buffer for incoming data, ensuring an exclusive ownership transfer that eliminates conflicts from multiple writers (either local or remote).
This built-in coordination removes the need for distributed locks or dedicated RDMA memory pools that incur copies, while avoiding data race conditions entirely. 
Our assessment in \S\ref{eval:rdma-primitives} shows that two-sided RDMA is \textbf{2\X-2.8\X} faster than one-sided write using distributed locks/Rendezvous, and up to \textbf{1.6\X} faster than one-sided write with receiver-side copy, which validates our choice.
The additional CPU overhead of two-sided RDMA (primarily from the receiver side) is often a concern~\cite{farm, prism-sosp21, citron}.
However, by delegating the two-sided RDMA operations to the DPU, we can preserve CPU cycles, while also maintaining the advantages of two-sided RDMA.

\vspace{-2mm}
\mybox{\textit{Design Implication\#3}: 
Two-sided RDMA is fundamentally more compatible with the requirements of a distributed, lock-free, zero-copy data plane in serverless environments, compared to one-sided RDMA.
}
\vspace{-2mm}

    \begin{figure}[t]
    \centering
        \includegraphics[width=\columnwidth]{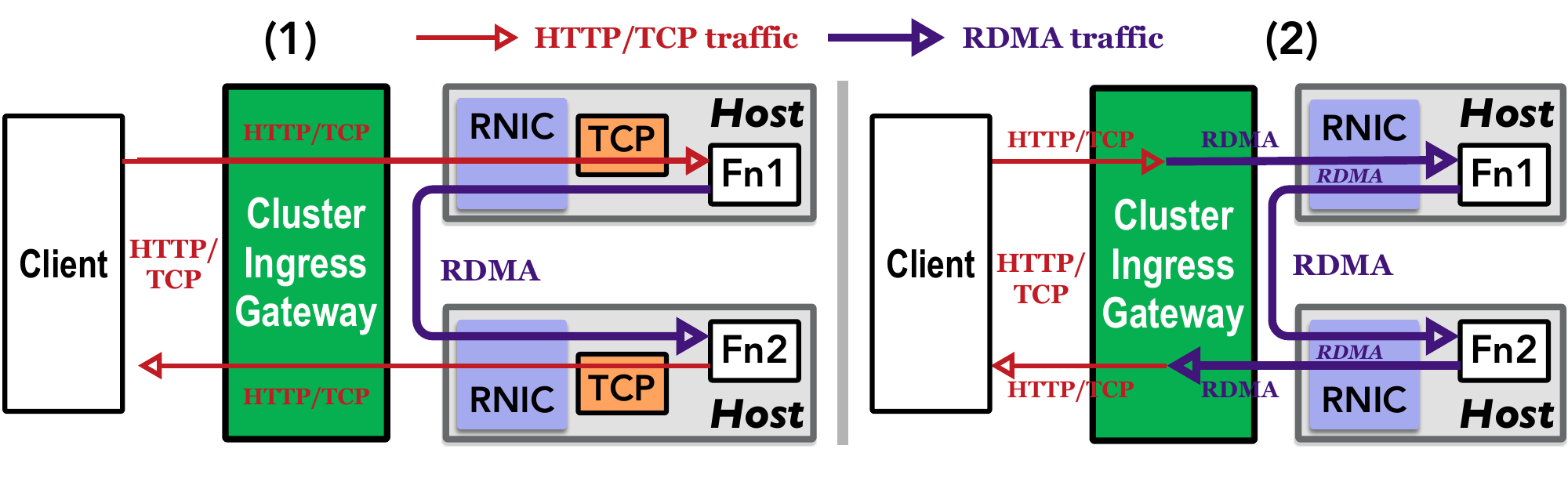}
    \vspace{-10mm}
    \caption{Alternatives of transport protocol adaptation: (1) ``Deferred'' transport conversion within cluster; (2) Early transport conversion at cluster ingress (used by \name).}
    \label{fig:transport-protocol-adapation}
    \vspace{-6mm}
    \end{figure}

\textnb{Challenge\#4: Transport Protocol Incompatibility.}
Fig.~\ref{fig:transport-protocol-adapation} (1) illustrates the transport mismatch described in \S\ref{sec:intro}: the existing RDMA-based serverless data plane relies on a TCP/IP stack on each worker node to terminate HTTP/TCP traffic from external clients. This ``deferred'' transport conversion results in duplicate HTTP/TCP/IP processing even within the serverless cluster.
Our evaluation in \S\ref{eval:cluster-ingress} reveals significant drawbacks in this approach. Specifically, the HTTP/TCP-based cluster ingress exhibits poor scalability, increasing the end-to-end latency by up to \textbf{11.7\X} and reducing the RPS by \textbf{11.4\X}.
An effective way to mitigate the redundant TCP/IP processing cost within the serverless cluster is to terminate the external client-side transport protocol at the \textit{earliest} point possible \ie at the cloud cluster edge, as shown in Fig.~\ref{fig:transport-protocol-adapation} (2).

\vspace{-2mm}
\mybox{\textit{Design Implication\#4}: 
Dedicating transport protocol adaptation to the cluster-wide ingress at the edge eliminates redundant TCP/IP processing within the cluster—an often-overlooked optimization that warrants more attention.
}
\vspace{-2mm}

\subsection{Related Work}\label{sec:related-work}

\textbf{Optimizing serverless data plane:}
SPRIGHT~\cite{spright}, NightCore~\cite{nightcore}, and many others~\cite{pheromone, faasm, yuanrong} rely only on local shared memory processing, which is not scalable.
While RMMap~\cite{rmmap} and FUYAO~\cite{fuyao} utilizes one-sided RDMA to construct their data plane, they are subject to the limitation of one-sided RDMA (\S\ref{sec:challenges}). Besides, none of them support multi-tenancy for RDMA and are limited by the traditional HTTP/TCP cluster ingress.
As summarized in Table-\ref{tab:compare-existing-designs}, \name goes beyond these prior designs with a comprehensively optimized data plane.
Other optimizations include lightweight sidecar implementations to reduce service mesh overhead~\cite{spright, sure, canal-mesh, vish-socc24}, and direct inter-function invocation~\cite{yuanrong, spright, fuyao, directfaas} to bypass the intermediate coordinator, thus completely avoiding the substantial cost imposed by such middleware.
These efforts complement \name.

\begin{table}[t]
\centering
\caption{Comparison of Existing High-Performance Serverless Dataplane Systems}
\vspace{-3mm}
\label{tab:compare-existing-designs}
\resizebox{\columnwidth}{!}{%
\begin{tabular}{lcccc}
\toprule
\multicolumn{1}{l}{\textit{Systems}} &
  \begin{tabular}[c]{@{}c@{}} \textbf{Multi-tenancy}\\ \textbf{Support}\end{tabular} &
  \begin{tabular}[c]{@{}c@{}} \textbf{Distributed}\\ \textbf{Zero-copy}\end{tabular} &
  \begin{tabular}[c]{@{}c@{}} \textbf{DPU}\\ \textbf{Offloading}\end{tabular} &
  \begin{tabular}[c]{@{}c@{}} \textbf{Eliminate}\\ \textbf{proto. processing}\\ \textbf{within cluster}\end{tabular} \\ 
\midrule
NightCore~\cite{nightcore} & \xmark & \xmark & \xmark & \xmark \\ 
SPRIGHT~\cite{spright}     & \xmark & \xmark & \xmark & \xmark \\ 
FUYAO~\cite{fuyao}         & \xmark & \xmark & \cmark & \xmark \\ 
RMMAP~\cite{rmmap}         & \xmark & \cmark & \xmark & \xmark \\ 
\name                      & \cmark & \cmark & \cmark & \cmark \\ 
\bottomrule
\end{tabular}%
}
\vspace{-6mm}
\end{table}

\textnb{RDMA:}
Several complementary efforts seek to improve various aspects of RDMA, including scalability~\cite{srnic, flock-sosp21, flor-osdi, irn}, 
and security~\cite{Bedrock}.
Performance isolation of RDMA has been a key focus in multi-tenant environments~\cite{collie, husky, harmonic, Justitia}.
Solutions like \cite{lite} and \cite{krcore} provide in-kernel mechanisms to isolate RDMA QPs from untrusted applications but require kernel changes.
\name enables the RDMA isolation through a userspace software solution (DNE), leveraging DPU offloading to eliminate CPU cost while also providing physical protection against CPU-based attacks.

\textnb{DPU/SmartNIC offloading}
has been widely explored to enhance various aspects of cloud data centers, including storage systems~\cite{BurstCBS, tianmen, Gimbal, alibaba-luna-solar-system}, distributed file systems~\cite{linefs}, multi-tenancy~\cite{OSMOSIS, panic}, request scheduling and load balancing~\cite{RingLeader, laconic, AlNiCo}, TCP offloading~\cite{flextoe, iotcp, AccelTCP}, service mesh acceleration~\cite{flatproxy}, distributed transactions~\cite{xenic}, ML training/serving~\cite{conspirator, splitrpc}. \cite{s-nic-minlan} provides isolation for SmartNIC-offloaded network functions from different tenants. Specific to serverless computing, \cite{lambda-NIC} offloads serverless functions to P4-enabled Netronome SmartNICs~\cite{netronome} but is constrained by a limited programming model. \cite{fuyao} uses DPU to offload the coordinator (Fig.~\ref{fig:serverless-data-plane-overview}), reducing CPU costs. \name goes beyond \cite{fuyao} by unleashing the potential of DPU offloading through cross-processor shared memory.

\section{Design of \name}

\subsection{Overview}
\begin{figure}[t]
    \centering
        \includegraphics[width=\columnwidth]{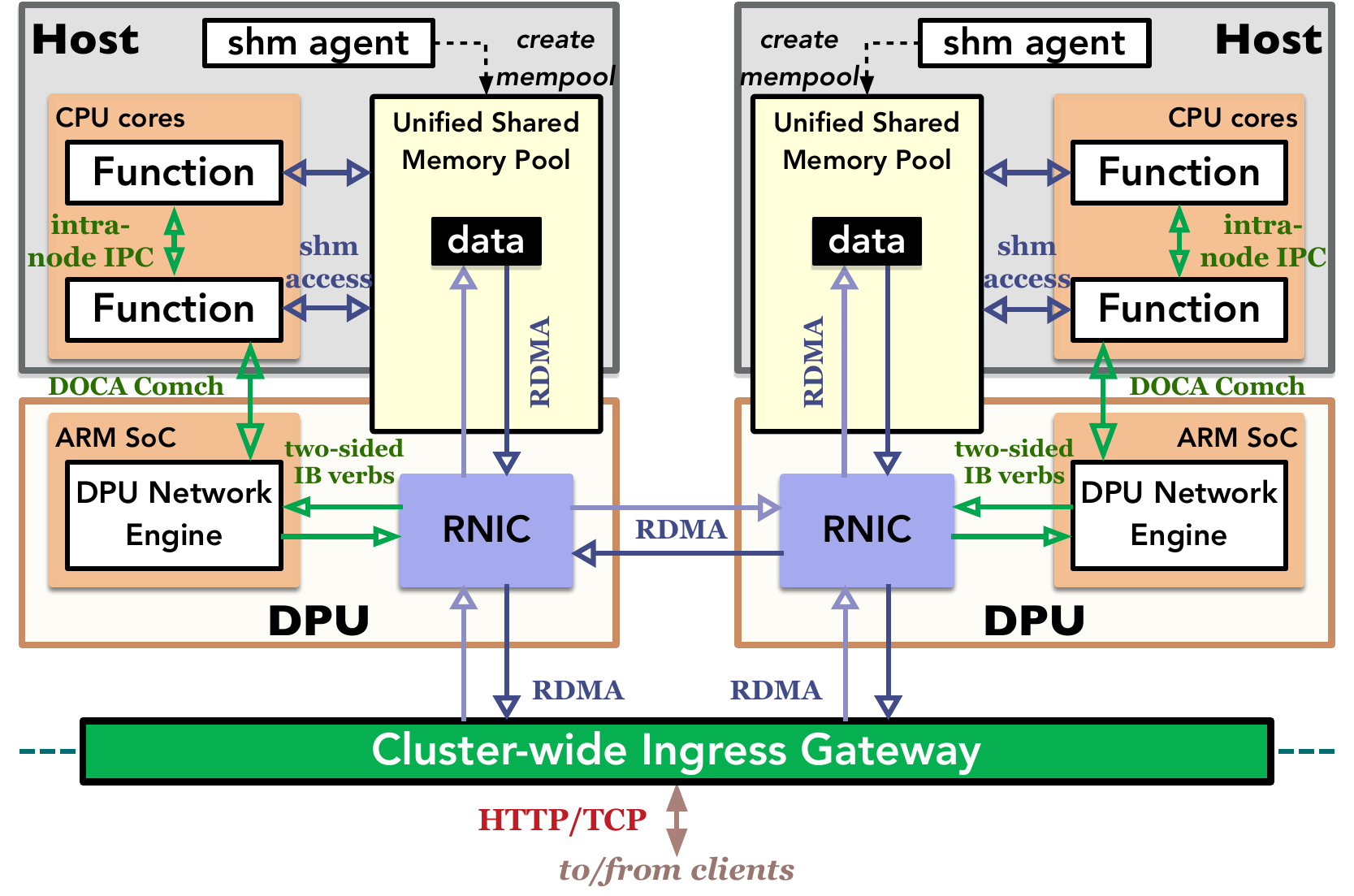}
    \vspace{-8mm}
    \caption{A architecture-level overview of \name.}
    \label{fig:palladium-overview}
    \vspace{-6mm}
    \end{figure}

Fig.~\ref{fig:palladium-overview} shows the core components in \name: 
\textnb{(1)}
Each worker node in \name has an associated DPU\footnote{Our implementation leverages the NVIDIA Bluefield-2~\cite{bluefield}, but our design is generally applicable to other Bluefield models.}, which  runs a lightweight DPU Network Engine (DNE) (\S\ref{sec:dne}).
By granting the DNE {\it exclusive} access to RDMA QPs, \name enforces strict per-tenant network isolation at the RNIC (\S\ref{sec:multi-tenancy}).
\textnb{(2)} \name treats each function chain as an independent `tenant', 
with each tenant owning a unified memory pool (physically located on the host) for exclusive access. This enables per-tenant memory isolation (\S\ref{sec:mem-isolation}).
By extending the unified memory pool into a cross-processor shared memory (\S\ref{sec:x-processor-shm}) between the DPU and the CPU (host), the DNE stitches the inter-node (RDMA) data plane into intra-node (shared memory) data plane, while being off-path.
\name employs the pool-based buffer management for fast buffer allocation and reuse.
\textnb{(3)} To facilitate lock-free communication in \name's data plane, we employ two-sided RDMA for inter-node data transfers and use token passing  for intra-node data transfers (\S\ref{sec:lock-free}).
On the host, we use the eBPF's \skmsg IPC from~\cite{spright} to hand off a buffer descriptor between co-located functions, taking advantage of its event-driven execution model and sidecar extensibility.
We rely on NVIDIA's DOCA Communication Channel (Comch~\cite{doca-comch}) for cross-processor exchange of buffer descriptors between the DNE and host functions, which was the most functionally capable option on Bluefield DPUs, as we evaluate in \S\ref{sec:x-processor-comch}.
\textnb{(4)} The ingress gateway is deployed on independent server nodes at the cluster edge, rather than on nodes acting as workers for serverless functions.
HTTP/TCP connections are terminated at the cluster-wide ingress gateway and the payload data is then transferred to nodes within the cluster over RDMA Fabrics, and vice versa (\S\ref{sec:cluster-ingress}).

\textnb{Security model and lightweight service mesh support.}
\name adopts a tenant-based security model~\cite{spright, rmmap} to restrict access to shared data: only functions from the same tenant that trust each other can use shared memory processing (see \S\ref{sec:shared-memory}). An explicit CPU-based data copy is used for communication across security domains.
We also utilize the service mesh to enforce the necessary access control when exchanging data between functions.

The existing service mesh deploys a separately-running sidecar (e.g., as a container) independent of the function, which results in additional network overhead~\cite{spright, mrpc, meshinsight, sure, vish-socc24}. This overhead can be as high as 30\% when based on the kernel network stack~\cite{spright}.
To reduce the overheads of service mesh, \name incorporates two optimizations to replace the heavyweight container-based sidecar:
(1) Leveraging consolidation and deploy a per-node ``shared'' sidecar~\cite{mrpc, vish-socc24, canal-mesh}, integrated with \name's DNE. This avoids duplicate sidecar processing during invocations between functions. 
(2) The second optimization uses a streamlined eBPF-based sidecar for each function~\cite{spright, cilium}, thereby avoiding the high container networking overhead. 
Both strategies support non-intrusive deployment and can isolate sidecar from (untrusted) user code.

\subsection{Off-path DPU Network Engine (DNE)}\label{sec:dne}

\name's DNE consists of a \textit{core thread} and multiple \textit{worker threads}. The core thread manages memory mapping from the host, registers memory regions to RNIC (\S\ref{sec:x-processor-shm}), and establishes DOCA Comch channels with host functions.
The worker thread handles the data packet processing 
within a non-blocking, {\it run-to-completion} event loop, processing each packet through all the transfer stages without interruption.
This {\it run-to-completion} design eliminates overheads (e.g., scheduling, context switch) that may tax the wimpy DPU cores, and further improves the cache locality~\cite{ix-osdi, demikernel}.

\textnb{Run-to-completion Event Loop:} 
Fig.~\ref{fig:datapath-details} describes the {\it run-to-completion} event loop, including both transmit (TX) and receive (RX) stages. 
In the TX stage, the DNE consumes a buffer descriptor from the source function, determines the destination node via the inter-node routing table,
and selects the least-congested RC connection. It then wraps the descriptor into an RDMA work request (WR) and posts it to the RNIC for transmission.
Upon data arrival, the RX stage polls the completion queue entries (CQEs) and retrieves the corresponding buffer descriptors via a receive buffer registry table lookup (more details in \S\ref{sec:datapath-design}). It then extracts the destination function ID, and forwards the descriptors to the function using the corresponding Comch endpoint.

\subsection{Multi-tenancy support in \name's data plane}\label{sec:multi-tenancy}
\name leverages the DNE to ensure fair sharing of 
RNIC bandwidth among co-located tenants.
Traffic from tenants of greater importance is prioritized using a Deficit Weighted Round Robin-like~\cite{deficit-rr} scheduler.
By leveraging tenant-specific weights, the DNE prioritizes functions accordingly, enabling more frequent transfers for higher-priority tenants compared to lower-priority ones.

\textnb{RC connection management:} 
\name uses RCQPs\footnote{We use ``RC-based connection'' and ``RCQP'' interchangeably.}  to establish dedicated, point-to-point reliable connections between peer nodes.
Each tenant is allowed to utilize multiple RCQPs (proxied by DNE), where each comprises a Send Queue (SQ) and a Receive Queue (RQ). To reduce the QP memory footprint, all of a tenant's RCQPs share a single RQ, which is exclusively posted with buffers from that tenant's private memory pool. This guarantees the RNIC delivers incoming data into the correct pool.
All RCQPs on a given node share a single Completion Queue (CQ).

Establishing RC connections between peer nodes incurs significant overhead, as the connection setup time is non-negligible (of the order of tens of milliseconds)~\cite{krcore, flor-osdi}. To mitigate this, we employ a node-specific management approach, where a pool of established connections (to a remote peer node) is managed by the DNE of this node. 
In order to maintain a large number of established RC connections with negligible overhead, we employ the ``shadow'' QP mechanism from~\cite{rogue-socc18} to categorize RCQPs in the pool into \textit{active} and \textit{inactive} QPs:
A RCQP is considered active when it has WRs queued; otherwise, it is inactive~\cite{Tassel-apnet23}. 
Inactive RCQPs consume \textit{no} RNIC resources~\cite{rogue-socc18, flor-osdi}. As such, we only need to limit the total number of active RCQPs per node to avoid cache thrashing on the RNIC. 
The DNE dynamically activates or deactivates RC connections in proportion to the load between node pairs without any cross-node QP state synchronization~\cite{rogue-socc18}.

\subsection{Memory Subsystem in \name}\label{sec:shared-memory}

Fig.~\ref{fig:memory-sharing-design} shows the architecture of memory subsystem in \name.
The unified shared memory pool includes a set of buffers created on the host. We use hugepage memory (2MB size each) to create buffers, which helps reduce the memory footprint of the Memory Translation Table on the RNIC cache~\cite{srnic}.
The unified memory pool is mapped to the DPU and registered as memory regions for the RNIC by DNE.

\textnb{Pool-based buffer allocation and recycling:}
\name utilizes the memory pool to reserve a number of equal-size buffers before they are actually used by a function. When the function needs a new buffer, it gets one from the memory pool instead of calling a malloc-like API (e.g., glibc) to dynamically allocate memory each time it is needed. 
This helps  reduce allocation latency, fragmentation, and prevent memory leaks~\cite{mallacc, RT-Mimalloc, TCMalloc-study}. We currently only support a fixed-sized memory pool. 
We use the DPDK memory allocator interface to allocate a new buffer (\texttt{rte\_mempool\_get()}) and to release an existing buffer to the pool (\texttt{rte\_mempool\_put()}).

\subsubsection{Memory isolation across tenants}\label{sec:mem-isolation}
\begin{figure}[t]
    \centering
        \includegraphics[width=\columnwidth]{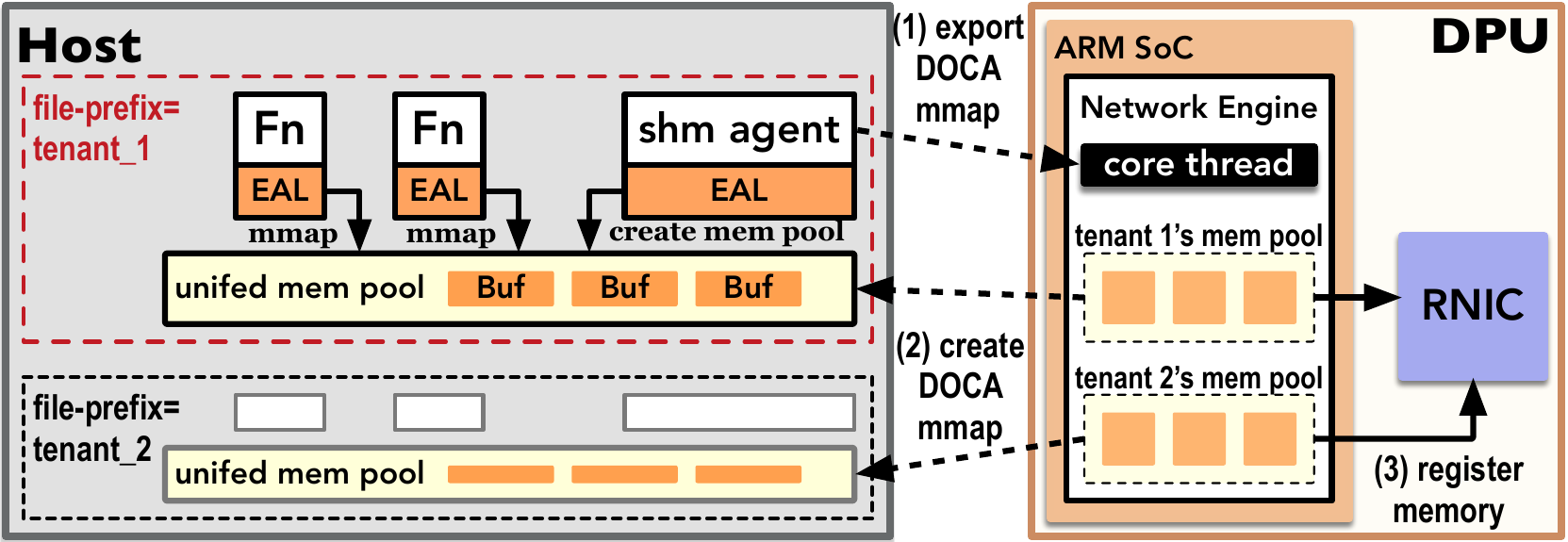}
        \vspace{-7mm}
    \caption{The memory sharing and isolation design in \name (a single node view). The memory pool is created on the host main memory and mapped to DPU.}
    \label{fig:memory-sharing-design}
    \vspace{-5mm}
    \end{figure}

\name uses the DPDK memory allocator to take advantage of DPDK's file-prefix feature~\cite{dpdk-multi_proc_support} to enforce per-tenant memory isolation, and generate memory-mapped files specific to the memory pool of a tenant (\ie a function chain).
The memory-mapped files contain the configuration (\eg the virtual addresses of the hugepages in use~\cite{dpdk-multi_proc_support}) of the memory pool created.
Each tanent has a distinct file-prefix, bound to a memory pool for exclusive access (as ``tenant\_1'' versus ``tenant\_2'' in Fig.~\ref{fig:memory-sharing-design}).

We create a per-tenant shared memory agent as the DPDK primary process, to set up the memory pool prior to the function startup (Note that the shared memory agent is not involved in the data transfer).
Functions (as DPDK secondary processes~\cite{dpdk-multi_proc_support}) utilize corresponding file-prefix to load the generated memory-mapped files and obtain the memory configuration. This enables functions to map to the shared memory pool through DPDK's Environment Abstraction Layer (EAL~\cite{dpdk-eal}) and
facilitates isolation between the memory pools of different tenants.

\subsubsection{Cross-processor shared memory}\label{sec:x-processor-shm}
\name achieves cross-processor shared memory by mapping the unified memory pool (physically located in the host's main memory) to the DPU. The cross-processor shared memory mechanism is implemented by using the mmap APIs from NVIDIA's DPU programming library (called DOCA~\cite{doca}). 

As shown in Fig.~\ref{fig:memory-sharing-design}, the DNE works in tandem with the shared memory agent to map the unified shared memory pool from the host:
(1) The shared memory agent on the host generates an export descriptor (DOCA's mmap descriptor output parameter which is used to represent memory ranges in remote system memory space) of the local memory pool. This is done using \texttt{doca\_mmap\_export\_pci()} (to grant DPU ARM core access) and \texttt{doca\_mmap\_export\_rdma()} (to grant RNIC access). The shared memory agent transmits it to the DNE via the DOCA Comch~\cite{doca-comch}.
(2) Upon receiving the export descriptor, the DNE establishes a remote memory map (using \texttt{doca\_mmap\_create\_from\_export()}). 
(3) With this setup complete, the DNE possesses all the necessary memory from the host, thus enabling the registration of the memory to the RNIC.

\subsection{Intra-node \& Inter-node data transfer}\label{sec:datapath-design}
Since \name's data plane spans both intra-node shared memory and inter-node RDMA, each uses distinct communication primitives.
To hide the intricacies 
of different data transfers from the user code, we introduce a unified I/O library built-in \name's function runtime. The I/O library offers unified send and receive APIs (\texttt{send()}, \texttt{recv()}) for operating on \name's data plane, sparing developers from selecting the correct transport.

\begin{figure}[t]
    \centering
        \includegraphics[width=\columnwidth]{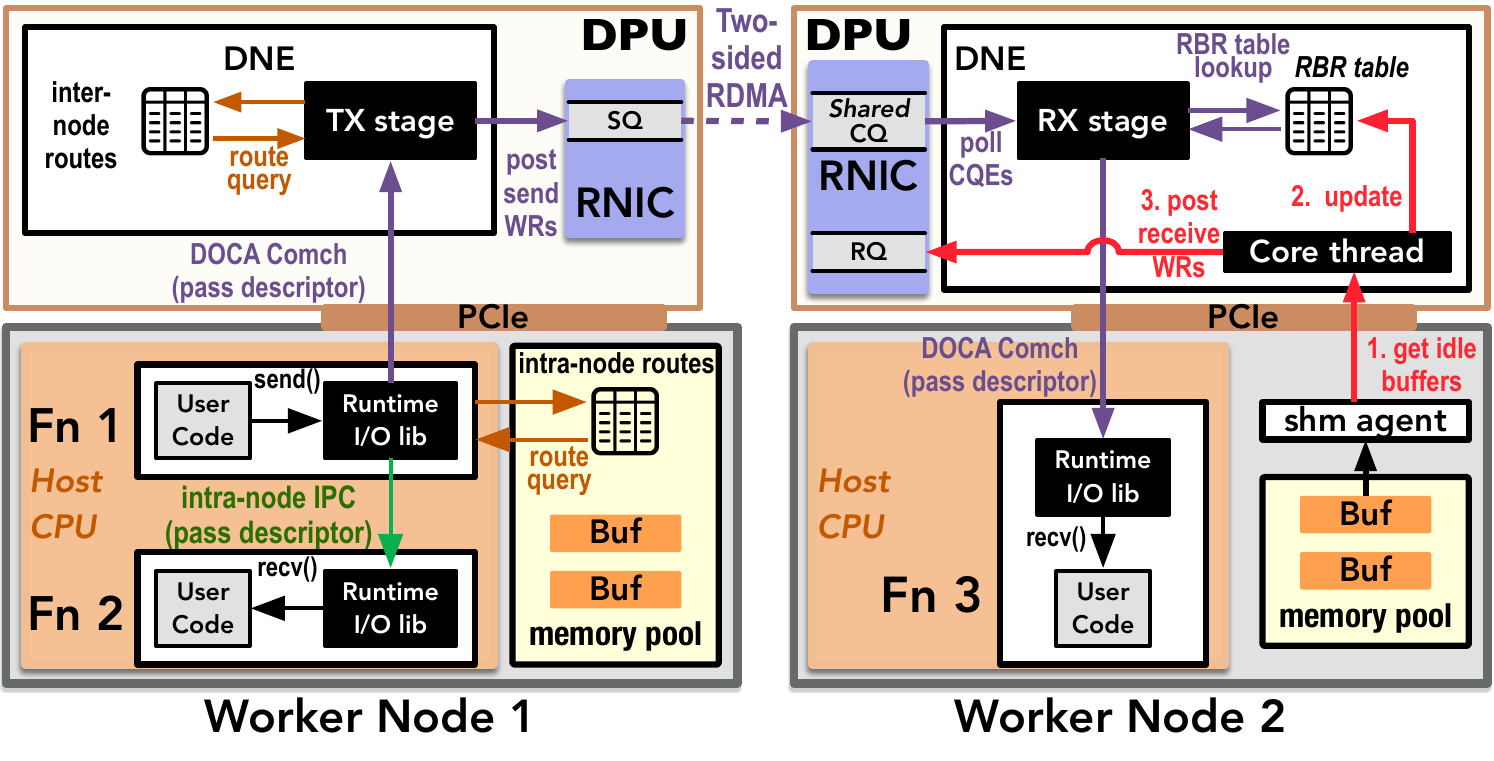}
        \vspace{-9mm}
    \caption{\textit{Lock-free, zero-copy} data transfer in \name. {\color{lincolngreen} Green arrow} depicts intra-node shared memory data transfer;
    {\color{violet} Violet arrows} depict inter-node data transfer using two-sided RDMA;
    {\color{red} Red arrows} depict RBR table updates (only shown on Node 2).}
    \label{fig:datapath-details}
    \vspace{-5mm}
    \end{figure}

The I/O library, once invoked by the user code, transparently determines the intra-/inter-node data path through the routing phases as shown in Fig.~\ref{fig:datapath-details}:
If the intra-node routing query confirms  that the destination function resides on the same node, the I/O library dispatches data using the intra-node shared memory IPC to exchange descriptors between functions ({\color{lincolngreen} green arrow} in Fig.~\ref{fig:datapath-details}).
Otherwise, it hands off the data to the DNE, which handles inter-node RDMA forwarding
({\color{violet} violet arrows} in Fig.~\ref{fig:datapath-details}).

\subsubsection{Lock-free buffer ownership transfer.}\label{sec:lock-free}
The use of two-sided RDMA prevents concurrent writes to the receiver-side buffer when inter-node data transfers occur.
In addition, the intra-node ownership transfer relies on explicit token passing of descriptors between local functions (or between host functions and DNE), which emulates the behavior of single-producer single-consumer ring that guarantees lock-free buffer access.
\name's buffer lifecycle management adheres to exclusive ownership semantics, \ie only the buffer owner (function or DNE) can read, write, or recycle the buffer, which is suitable for working with the message-passing communication model using two-sided RDMA.
This helps ensure that buffers are not inadvertently modified (data corruption) or prematurely released (may cause segmentation faults).

\subsubsection{Inter-node RDMA transfers.}
Our two-sided RDMA design has the receiver-side DNE post a buffer (included in the WR) to the RNIC for receiving data from the sender. To track the buffer where the data is RDMA'ed into, we maintain a receive buffer registry (RBR) table on the DNE to map the WR to the posted receive buffer (similar functionality is already integrated when using DOCA APIs).
The DNE core thread asynchronously monitors the number of CQE 
consumed by different tenants (via shared counters updated by the RX stage) and posts an equal amount of receive buffers to the RQ of corresponding tenant.
This ensures the receiving RNIC always has enough buffers to receive incoming data for corresponding tenants
(shown as {\color{red} red arrows} in Fig.~\ref{fig:datapath-details}).

    \begin{figure}[b]
    \vspace{-4mm}
    \centering
        \includegraphics[width=\columnwidth]{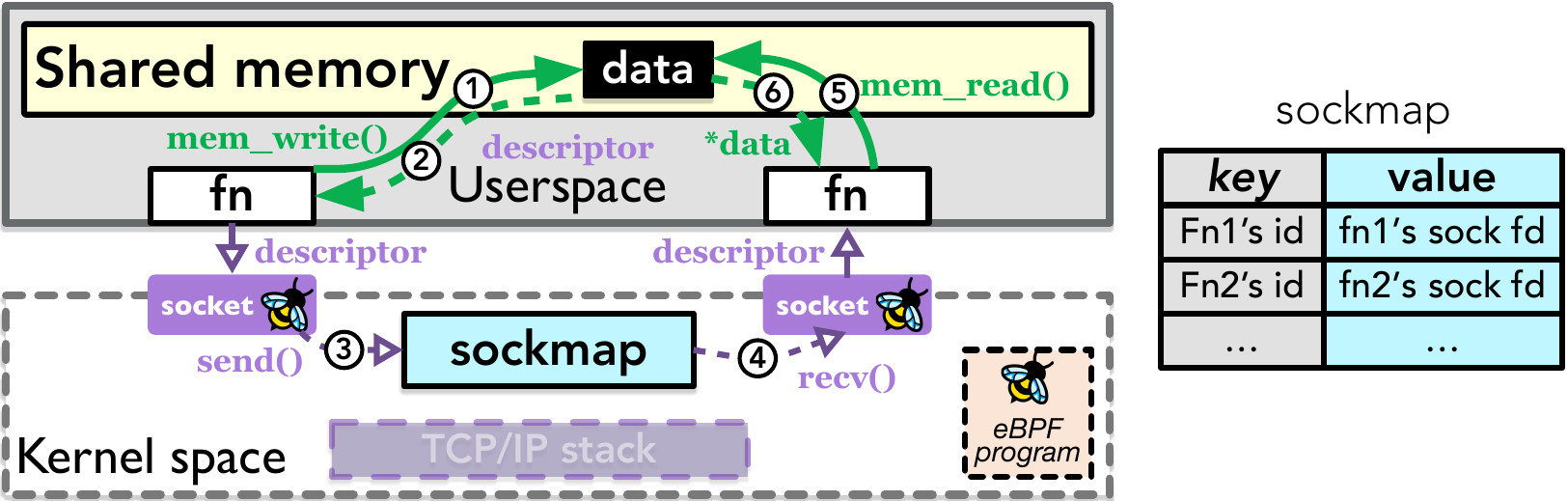}
        \vspace{-6mm}
    \caption{Intra-node (shared memory) data plane in \name.}
    \label{fig:intra-node-dp}
    \end{figure}

\subsubsection{Intra-node shared memory IPC}
The zero-copy shared memory processing between functions depends on delivering the descriptor of the shared memory buffer.
The ownership of the descriptor determines the access to the shared memory buffer, using the data pointer contained in the descriptor.
\name utilizes eBPF's \skmsg~\cite{skmsg} for intra-node IPC (same as~\cite{spright}).
The transmission of \skmsg between sockets entirely bypasses the kernel protocol stack, thus eliminating unnecessary processing, and making it a desirable capability to transfer the small descriptors.

Fig.~\ref{fig:intra-node-dp} depicts the intra-node data flow between two \name functions. The source function uses POSIX \texttt{send()} (wrapped in the I/O library's unified send API) to transmit the descriptor, which triggers the execution of \skmsg. The \skmsg look up the \textit{sockmap} to find the socket interface of the destination function so that the descriptor is delivered to it for access of the shared memory buffer. The \textit{sockmap} is a special eBPF map (\texttt{BPF\_MAP\_TYPE\_SOCKMAP}~\cite{sockmap}) that keeps references to registered socket interfaces.

\subsubsection{Cross-processor communication channel.}\label{sec:x-processor-comch}
We deploy the DNE as the single Comch server instance to communicate with functions (which are Comch clients) on the host.
The DNE busy-polls all monitored function endpoints within its event loop for buffer descriptors. We use epoll to enable event-triggered reception by the function, ensuring a more efficient and practical approach that retains good function density in a serverless environment.

\begin{figure}[t]
    \centering
    \includegraphics[width=.5\columnwidth]{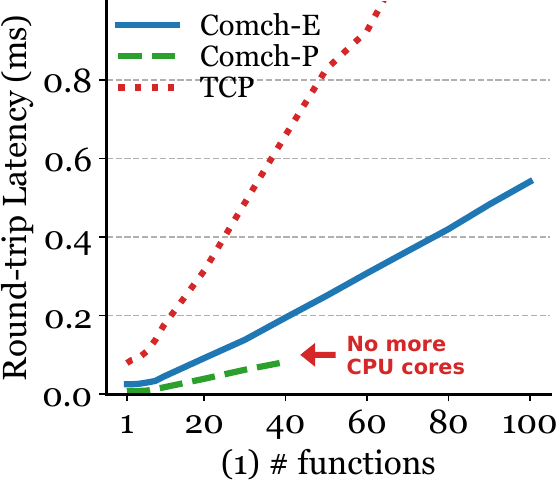}\hfill
    \includegraphics[width=.5\columnwidth]{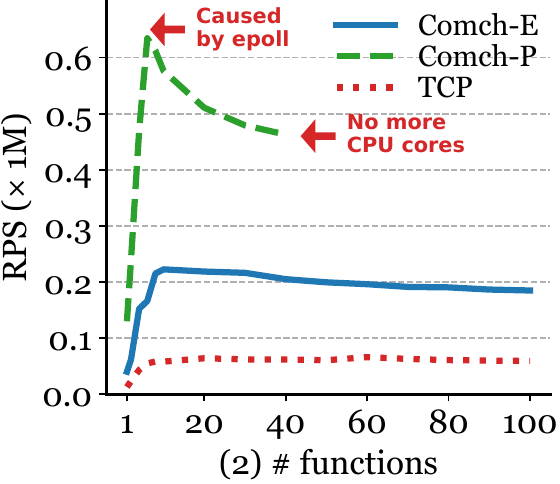}
    \vspace{-8mm}
    \caption{Evaluation of viable communication channels between DPU and host (CPU): (1) Round-trip latency (in milliseconds); (2) Buffer descriptor transfer rate (RPS).}
    \vspace{-5mm}
    \label{res:x-proc-comch}
\end{figure}

{DOCA Comch offers two communication channel variants~\cite{doca-comch}, which can be used for exchanging 16B buffer descriptors between \name's DNE and host functions: (1) Comch-E uses event‐driven send/receive primitives on top of blocking Linux epoll, while (2) Comch-P uses a producer–consumer ring with busy polling (lowest latency but ties up a core per function). We compare both against a TCP baseline (we avoid intra-node RDMA as it is insecure, allowing user functions direct access to QPs violating multi-tenancy, as discussed in \S\ref{sec:challenges}. In contrast, Comch allows the DNE to disconnect misbehaving tenants). In our test, multiple functions on the host issue back‐to‐back descriptor sends to a single‐core DNE on the DPU and await replies.

Fig. 8 shows the comparison results: TCP suffers the highest latency due to kernel and protocol overhead. Comch-P cuts latency by >8\X versus TCP but hits scalability limits (one core per function) and overloads beyond 6 functions.
Based on our examination, it is caused by the use of an internal epoll interface by \texttt{Comch-P} for its ``busy'' polling (DOCA implements its event loop called ``Progress Engine''~\cite{doca-pe}, which is actually performed via non-blocking \texttt{epoll\_wait} and introduces kernel-related overheads, which hopefully will be improved upon in the future in this closed-source implementation).
Comch-E, while is slower than Comch-P, still outperforms TCP by 2.7\X–3.8\X and delivers stable. Its event‐driven design without dedicated cores makes it the most practical choice for a dense, multi‐tenant serverless platform. We choose to use Comch-E in \name.

\subsubsection{Routing state management}
The intra-node routing table (on the host) maintains the routes between local functions. We store the intra-node routing table in the unified memory pool as shared states and make it read-only for the functions.
The inter-node routing table (on the DPU) maintains the routes to the remote functions. 
We have a coordinator in the control plane (similar to the Container Network Interface (CNI)~\cite{cni}) listens for function deployment events (creation, termination) and synchronizes the up-to-date intra-node routes and inter-node routes on the worker nodes.

\subsection{\name's Cluster-wide Ingress Gateway}\label{sec:cluster-ingress}
The ingress gateway employs a master-worker model, as shown in Fig.~\ref{fig:ingress}. The master process manages control-plane tasks, including loading the configuration and horizontal scaling of the gateway worker processes. Worker processes handle all data-plane tasks, including TCP/IP protocol processing, HTTP processing, and RDMA transmission. Each worker process performs all data-plane tasks for transport protocol adaptation, using a run-to-completion event loop (busy polling loop), just like the DNE (\S\ref{sec:dne}). We enable batching in the event loop to improve concurrency.
We base the implementation of the ingress on NGINX~\cite{nginx} (v1.25.2) to leverage its full-fledged HTTP processing implementation. We integrate the DPDK-based F-stack~\cite{f-stack} in our ingress's busy polling loop taking advantage of its high-performance TCP/IP protocol processing in the userspace instead of the slower kernel protocol stack.
Deploying \name's cluster gateway does not require a DPU. A node equipped with a regular RNIC and a DPDK-compatible NIC is sufficient.

    \begin{figure}[b]
    \vspace{-5mm}
    \centering
        \includegraphics[width=\columnwidth]{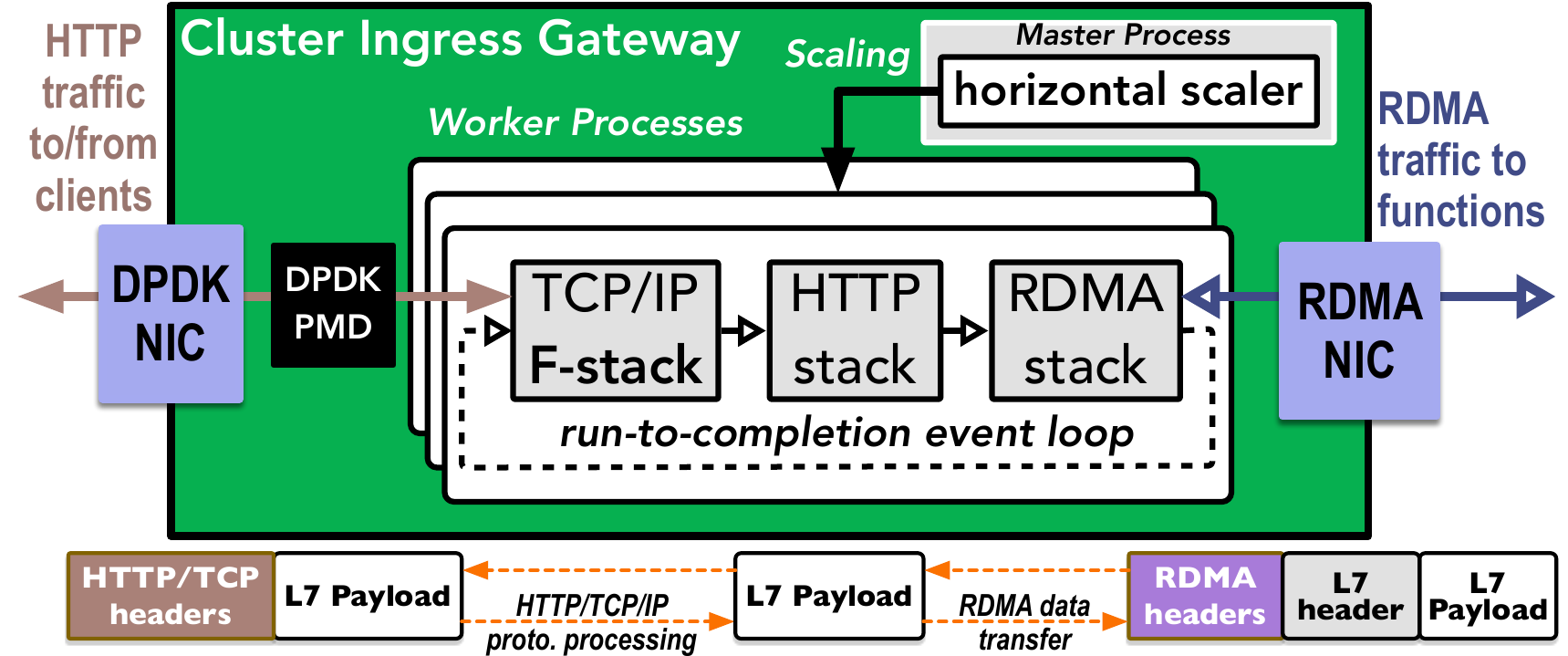}
        \vspace{-7mm}
    \caption{HTTP/TCP-to-RDMA transport protocol adaptation at the cluster-wide ingress gateway.}
    \label{fig:ingress}
    \end{figure}

\textnb{Event-driven architecture:}
Callback functions are registered in the event loop to handle HTTP protocol processing (as in NGINX~\cite{nginx}) and RDMA send/receive events. 
When an event occurs (\eg on a new connection or data is available to read), the corresponding handler is called to process it.
We leverage Receive Side Scaling (RSS~\cite{qizhe-21}) to distribute traffic from external clients evenly to different worker processes (pinned to specific CPU cores), thus achieving the equivalent effect of Accelerated Receive Flow Steering (aRFS) without additional NIC hardware support~\cite{qizhe-21}.

\textnb{Horizontal scaling:}
We allocate each worker process a dedicated CPU core to execute the busy polling loop. 
To optimize resource usage during low demand periods and ensure performance during peak traffic, we horizontally scale the number of worker processes based on the load level of the worker processes using a simple but effective hysteresis policy. 
We refine the CPU time measurement in the worker process event loop to collect the aggregated CPU time spent on data-plane tasks. This enables us to quantify  CPU utilization dedicated to the ``useful'' work performed by the worker process.
Once the average CPU utilization across existing worker processes reaches 60\%, the master process spawns a new worker process to handle the increased load. Conversely, when the average CPU utilization across existing worker processes drops below 30\%, the master process terminates a worker process to conserve CPU resources.

\section{Evaluation}

We first perform a set of microbenchmarks to showcase the improvement by the various design choices made in \name. We then show the benefit of \name with a system-level evaluation using real serverless workloads.

\textnb{Testbed:}
Our testbed consists of four server nodes. Each node is equipped with two 40-core CPUs (NUMA enabled) and 500GB of RAM. We run Ubuntu 22.04 with kernel 5.15 on each node. Two of the nodes are used as workers for deploying user functions. Each worker node has an NVIDIA Bluefield-2 DPU (with an integrated ConnectX-6 RNIC) to act as the DNE. We use another node (called the ingress node) to deploy the cluster-wide ingress gateway and the remaining node without the DPU as the client node to generate the workload from external clients.
The ingress node has two separate ConnectX-6 RNICs: one used for RDMA communication with user functions and the other to function as the Ethernet NIC to communicate with the client node.
The DPUs on the worker nodes and the ConnectX-6 RNIC on the ingress node are interconnected by a 200 Gbps switch. The Ethernet NIC on the ingress node and the client node are connected through another distinct 200 Gbps switch.

\begin{figure}[t]
    \centering
    \includegraphics[width=.5\columnwidth]{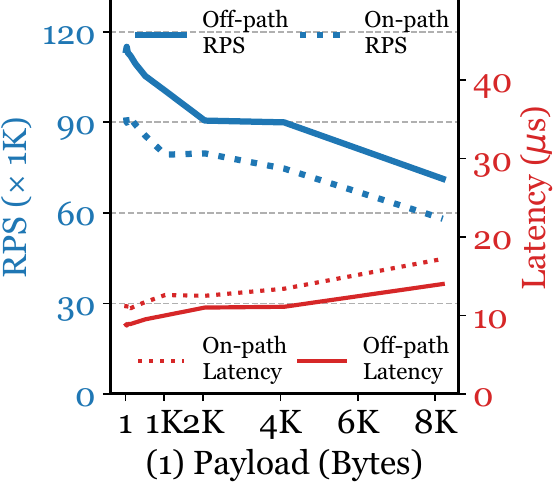}\hfill
    \includegraphics[width=.5\columnwidth]{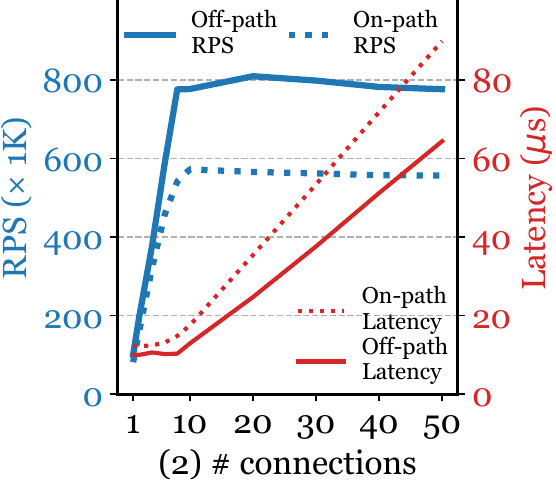}
    \vspace{-8mm}
    \caption{\name's off-path DNE (enhanced with cross-processor shared memory) VS. on-path DNE: (1) RPS with varying payload sizes (single connection); (2) RPS under different concurrency levels (1KB payload size).}
    \label{res:x-proc-shm}
    \vspace{-4mm}
\end{figure}

\subsection{Microbenchmark Analysis}

\subsubsection{Cross-processor shared memory}\label{eval:x-processor-shm}
We compare the performance difference between the offloading model with off-path DPU processing in \name (enhanced with cross-processor shared memory) versus offloading with on-path DPU processing. The two options are depicted in Fig.~\ref{fig:x-processor-shm}. In the on-path mode, the DNE buffers data locally (SoC memory) when moving the data between a function on the host and the RNIC. The on-path DNE uses IB verbs to interact with the RNIC, while using the DMA engine on DPU SoC to move data between local buffers and the shared memory pool on the host.
Note that we do not use intra-node RDMA to move data between DPU and the host as it adds 2 PCIe transfers~\cite{xingda}. On the other hand, RNIC DMA avoids this added overhead. 
It also avoids interference with inter-node RDMA traffic~\cite{mrpc}.
We use a echo server/client function pair to measure the round-trip time and throughput. The server and client functions are deployed on different nodes.

Fig.~\ref{res:x-proc-shm} shows the RPS and latency comparison: 
The off-path DNE achieves up to 30\% RPS improvement with more than 20\% latency reduction compared to the on-path mode under various payload sizes and concurrency levels. The shortcoming of on-path mode is manifested under high concurrency traffic due to the poor processing capability of the DPU DMA engine~\cite{xingda}. 
This shows the necessity of using cross-processor shared memory between the host and DPU, thereby eliminating the need for additional data copies (and using DMA instead).
At low concurrency, the RPS of on-path mode is close to off-path mode because the DPU's DMA engine is not overloaded. At this light load, the on-path mode can fully exploit the low-latency data copy of the DPU's DMA engine (only 2.6$\mu$s for 64B DMA read~\cite{xingda}). But as the number of connections increases, the on-path mode's latency grows much more quickly compared to the off-path mode.

\subsubsection{Selection of RDMA primitives}\label{eval:rdma-primitives}
To validate our choice of two-sided RDMA, we benchmark three alternatives in Fig.~\ref{fig:RDMA-primitives-selection}. We configure two DNEs on different worker nodes
to act as a pair of echo client/server and allocated one core to each.  
Since the completion of the one-sided RDMA write is not visible to the receiver, it has to poll for the data arrival (following FARM's receiver-side design~\cite{farm}).
The receiver then echos the data.
Note that repetitive measurements with OWRC can introduce bias on the echo server. Specifically, the source buffer in the RDMA memory pool and the destination buffer in the local memory pool may both be cached.
This can lead to optimal cache locality and an artificially improved copy performance. This scenario is unlikely to hold in real-world workloads. To account for this, we introduce a variant of OWRC that enforces main memory access during data copies by flushing the TLB, ensuring a worst-case evaluation. We refer to the OWRC with artificial cache locality as \texttt{OWRC-Best} and the one with enforced main memory access as \texttt{OWRC-Worst}.

\begin{figure}[t]
    \centering
    \includegraphics[width=.5\linewidth]{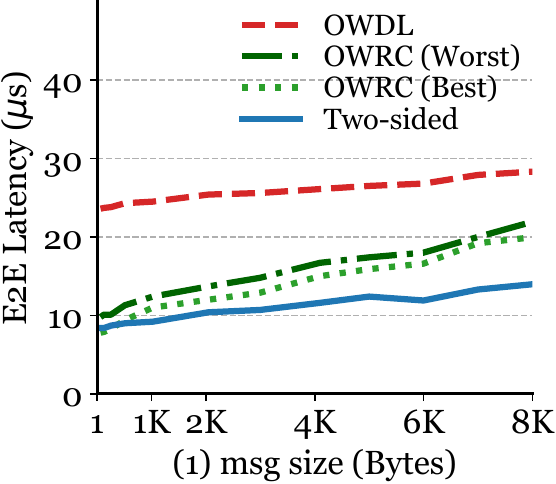}\hfill
    \includegraphics[width=.5\linewidth]{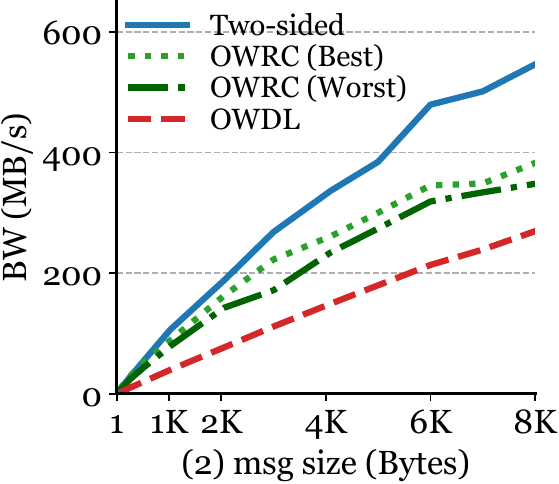}
    \vspace{-8mm}
    \caption{Performance impact of selecting RDMA primitives: (1) mean end-to-end latencies; (2) RPS. \name uses two-sided RDMA.}
    \vspace{-5mm}
    \label{res:rdma-primitives-perf}
\end{figure}

Fig.~\ref{res:rdma-primitives-perf} (1) shows the latency differences between the various approaches. A single one-sided write can take as little as 4$\mu$s, but exchanging locks and copies introduce non-negligible overhead.
Two-sided RDMA achieves the lowest latency (8.4$\mu$s for 64B  messages) by eliminating data races in the data plane (\S\ref{sec:challenges}), thereby streamlining data transfers over RDMA.
With a 4KB message size, two-sided RDMA (11.6$\mu$s) reduces latency by 1.3\X, 1.5\X and 2.3\X compared to \texttt{OWRC-Best} (15$\mu$s), \texttt{OWRC-Worst} (16.7$\mu$s) and \texttt{OWDL} (26.1$\mu$s), respectively.
Looking at throughput in Fig.~\ref{res:rdma-primitives-perf} (2), \name is up to 1.3\X and 1.4\X higher than \texttt{OWRC-Best} and \texttt{OWRC-Worst}, and is more than 2.1\X more than \texttt{OWDL}.
These demonstrate the advantages of adopting two-sided RDMA in \name.

\subsubsection{Transport protocol adaptation}\label{eval:cluster-ingress}
We evaluate end-to-end latency savings and improved scalability from consolidating the transport protocol adaptation to \name's cluster edge (Fig.~\ref{fig:transport-protocol-adapation} (2)). 
As baselines, we compare two NGINX‐based ingresses
The baselines we compared with are two NGINX‐based variants of Fig.~\ref{fig:transport-protocol-adapation} (1): one uses the interrupt-driven Linux kernel TCP/IP stack (denoted as \textit{K-Ingress});
the other uses the more performant DPDK-based F-stack~\cite{f-stack} for protocol processing tasks (denoted as \textit{F-Ingress}).
We deploy an HTTP server function on the worker node to echo the requests from ``external'' clients (wrk~\cite{wrk}), relayed by the cluster ingress. 
For the variants of Fig.~\ref{fig:transport-protocol-adapation} (1), we use F-stack on the worker node for TCP/IP processing.

\begin{figure}[t]
    \centering
    \includegraphics[width=.5\linewidth]{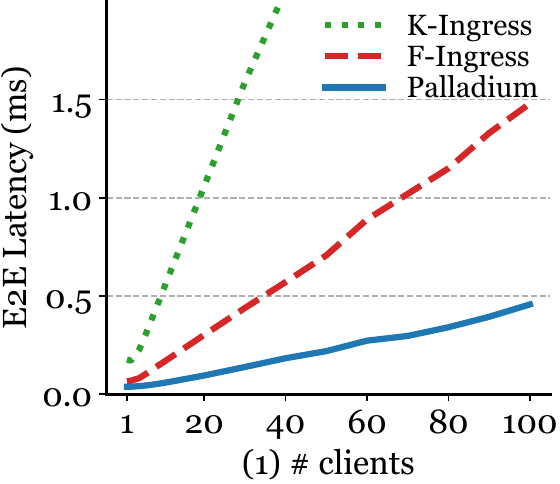}\hfill
    \includegraphics[width=.5\linewidth]{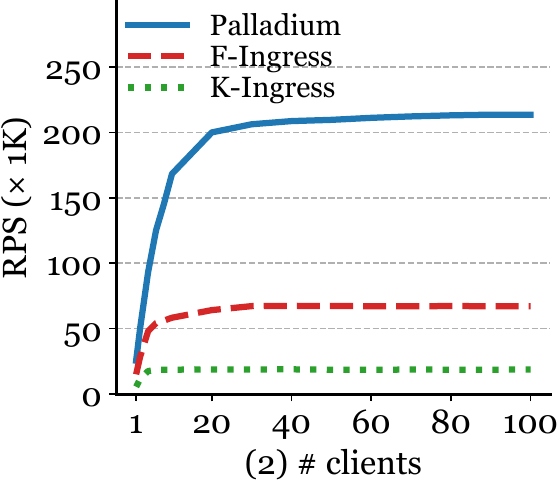}
    \vspace{-8mm}
    \caption{Performance of various cluster ingress designs: (1) mean end-to-end latencies; (2) RPS with varied \# of clients.}
    \vspace{-5mm}
    \label{res:cluster-ingress-perf}
\end{figure}

We assign one CPU core to the cluster ingress in this experiment.
The cluster ingress is likely to have to consistently handle a high volume of requests across concurrent connections. 
As shown in Fig.~\ref{res:cluster-ingress-perf} (1),
the end-to-end latency of \name's ingress is significantly lower than \textit{K-Ingress} and \textit{F-Ingress}, since \name replaces the slow software-based transport processing with RDMA for intra-cluster networking. \name also increases RPS by up to 11.4\X and 3.2\X RPS compared to \textit{K-Ingress} and \textit{F-Ingress}, respectively.
Since the variants of Fig.~\ref{fig:transport-protocol-adapation} (1) terminate the TCP at the worker node, this in fact doubles TCP/IP processing work at the cluster ingress, thus affecting scalability.

\begin{figure}[b]
    \vspace{-4mm}
    \centering
    \includegraphics[width=.5\linewidth]{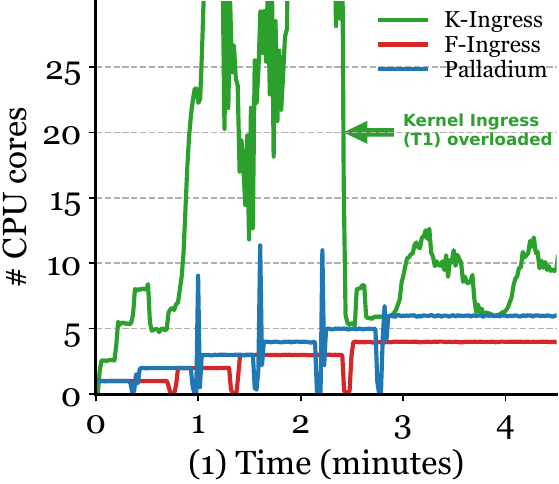}\hfill
    \includegraphics[width=.5\linewidth]{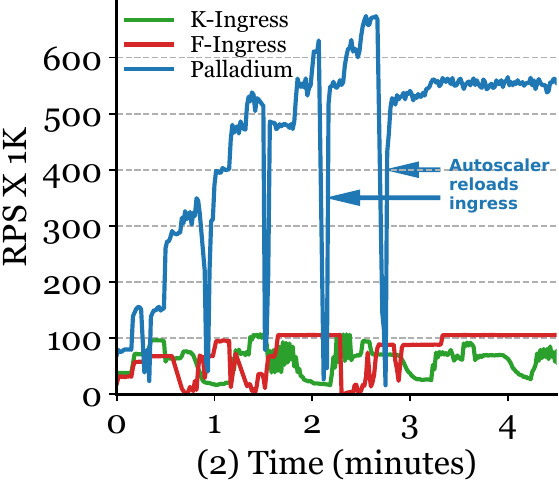}
    \vspace{-8mm}
    \caption{Effect of horizontal scaling of \name's ingress: (1) CPU usage time series of the cluster ingress; (2) RPS times series.}
    \label{res:cluster-ingress-hpa}
\end{figure}

To assess the horizontal scaling of \name's ingress, we increase the load by adding a client every 10 seconds, and each client is configured to fully use up a CPU core to generate the highest load it can (with multiple connections). We also adapt our autoscaler to support the \textit{F-Ingress} for a fair comparison.
Fig.~\ref{res:cluster-ingress-hpa} (1) shows the CPU usage over time.
The horizontal scaling in \name's ingress alleviates the busy-polling overhead by adjusting the number of active worker processes to match load, while retaining the low-latency benefit of busy polling (see \S\ref{sec:cluster-ingress}). 
\name's ingress uses much less CPU while achieving more than 5\X RPS compared to interrupted-driven \textit{K-Ingress}. The \textit{K-Ingress} is quickly overloaded after using up all CPU cores (at around the 2.5 minute mark) and results in most of the clients becoming disconnected due to the lack of a response.
Note that the scaling procedure in \name's ingress triggers a brief service interruption due to the restart of the worker processes (see Fig.~\ref{res:cluster-ingress-hpa} (2)). This can be avoided by enabling load balancing across multiple \name ingress instances.

\subsection{Multi-tenancy Support for RDMA}\label{eval:multi-tenancy}
\textnb{Tenant types and workload:}
We assess the effect of multi-tenancy support in \name (via the implemented DWRR-like policy in DNE) for the fair sharing of inter-node bandwidth offered via DNE. We compare \name 
with a first-come-first-served (``FCFS'') DNE that does not explicitly perform functions to support multi-tenancy. 
In this experiment, we configure the DNE to sustain a maximum throughput of approximately 110K RPS when it fully utilizes the assigned single DPU core.
We consider three tenants that compete for the bandwidth, each assigned a different weight: {Tenant-1}'s weight is 6; {Tenant-2}'s weight is 1; {Tenant-3}'s weight is 2.
Every tenant runs a pair of client/server functions, placed across two worker nodes to trigger inter-node data transfers via the DNE. 
We induce contention by making Tenant-2 and Tenant-3 generate periodic surges (Tenant-3  is slightly more bursty).
{Tenant-1} (green line) remains active for the entire 4-minute experiment, while {Tenant-2} (dotted line) joins at 20s and exits at 3m20s. {Tenant-3} runs between 1m30s and 2m30s (blue line).

\textnb{Observations:}
Fig.~\ref{res:multi-tenancy-support} shows the bandwidth distribution among tenants, using (1) the `FCFS' DNE and (2) \name's DNE with multi-tenancy support. 
The FCFS DNE simply processes requests in order, favoring tenants that send requests earlier rather than distributing bandwidth according to predefined weights.
This leads to \textit{unfair} bandwidth sharing, as evidenced by the starvation effect observed for {Tenant-1} once additional bursty tenants join. 
In contrast, \name's DNE shows improved fairness
by precisely scheduling traffic based on their allocated weights when different tenants are competing for bandwidth. All received a weighted fair share in a controlled manner: Upon {Tenant-2}'s arrival, it received 15K RPS, while {Tenant-1}'s RPS decreased from 115K to 90K - precisely maintaining the 1:6 ratio based on their weights. When {Tenant-3} joined, the bandwidth distribution adjusts to 65K, 11K, and 22K for Tenant-1, 2, and 3 respectively, again aligning exactly with their weights. Tenant 1 is consistently processed at a higher rate.
Additionally, since \name supports multi-tenancy via a userspace software solution, it is easy for users to apply workload-specific optimizations by customizing policies in DNE.

\begin{figure}[t]
    \centering
    \includegraphics[width=.5\columnwidth]{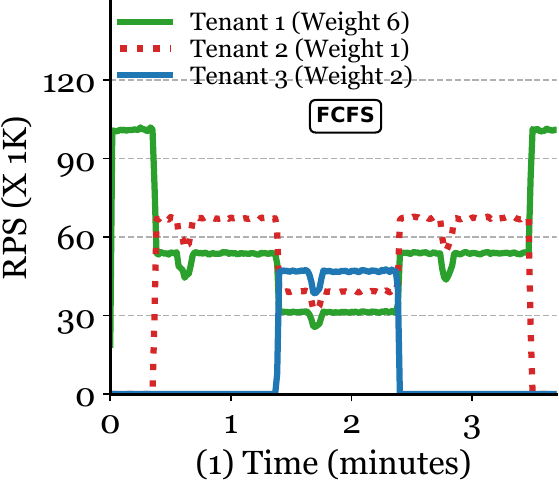}\hfill
    \includegraphics[width=.5\columnwidth]{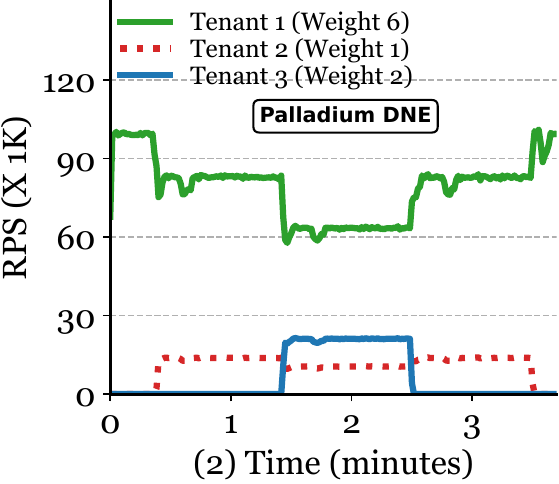}
    \vspace{-6mm}
    \caption{Effect of RDMA network isolation in \name: 
    (1) ``FCFS'' DNE {\it without} multi-tenancy support; (2) \name's DNE {\it with} multi-tenancy support.}
    \vspace{-5mm}
    \label{res:multi-tenancy-support}
\end{figure}

\begin{figure*}[t]
    \centering
        \includegraphics[width=\columnwidth]{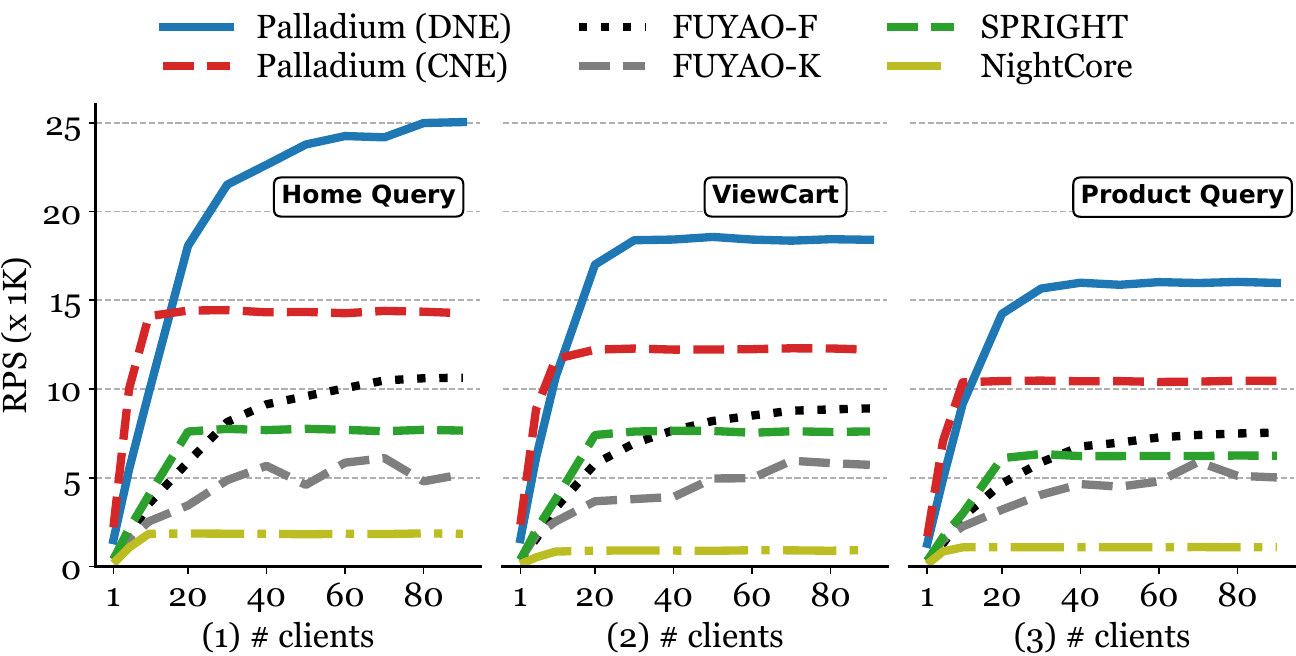}\hfill
        \includegraphics[width=\columnwidth]{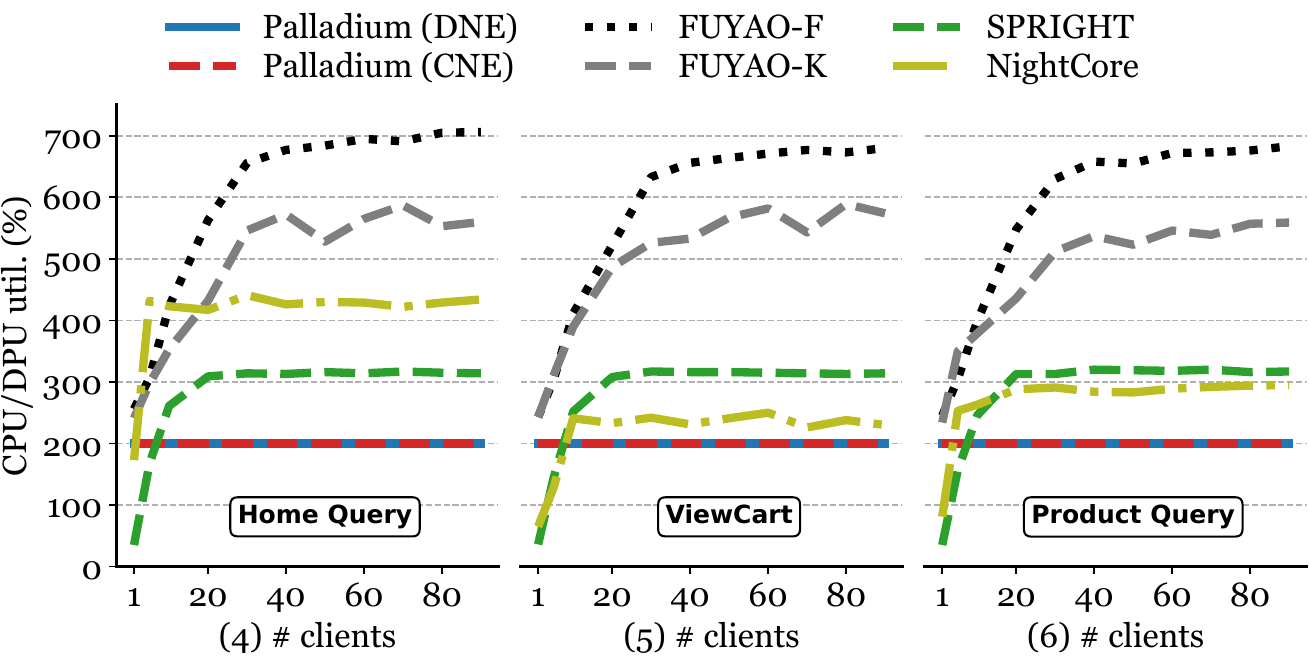}
    \vspace{-3mm}
    \caption{Online boutique results for different chains: Fig. (1) to (3): RPS; Fig. (4) to (6): Efficiency of offloading (CPU and DPU utilization). \name (DNE) shows DPU utilization. Others show CPU utilization.}
    \label{res:online-boutique-dpu-efficiency}
    \vspace{-4mm}
\end{figure*}

\subsection{Putting It All Together}\label{sec:real-workload}
\textnb{Real Workloads and Comparison with Baselines:}
We use the representative Online Boutique~\cite{online-boutique}, a microservices-based application, including 10 functions and offers up to 6 different function chains. Refer to~\cite{online-boutique} for the function call graph chains of the Online Boutique.
We focus on a distributed setup with two worker nodes to holistically evaluate the benefits of seamlessly integrating inter-node RDMA transfers with intra-node shared memory in \name: We place the potential hotspot functions (Frontend, Checkout, and Recommendation) on one node while placing the remaining functions on a second node.
We use the wrk~\cite{wrk} as the load generator.

\textnb{Data plane configurations:}
We compare \name with three advanced serverless data plane designs: SPRIGHT~\cite{spright}, NightCore~\cite{nightcore} and FUYAO~\cite{fuyao}.
All these approaches use local shared memory processing. As NightCore lacks support for inter-function communication across nodes within a function chain, we configure it to run all functions on a single node. SPRIGHT supports inter-node communication but relies on the kernel protocol stack, while FUYAO uses RDMA for inter-node communication but relies on one-sided writes with receiver-side copying. 
Similar to \name, each of these approaches incorporates a node-wide network engine-like component to facilitate data movement in and out of the local memory pool. 

Due to the severely poor overload behavior of kernel-based cluster ingress (\textit{K-Ingress}) identified in \S\ref{eval:cluster-ingress}, we use the F-stack NGINX (\textit{F-Ingress}) as the HTTP/TCP cluster ingress for SPRIGHT.
We additionally evaluate FUYAO with both \textit{K-Ingress} (denoted as ``\texttt{FUYAO-K}'')  and \textit{F-Ingress} (denoted as ``\texttt{FUYAO-F}'') in \S\ref{eval:cluster-ingress}.
NightCore relies on its built-in kernel-based cluster ingress.
To quantify the benefit of DPU offloading, we additionally run \name's Network Engine on the CPU (version called ``\name(CNE)'') for an apples-to-apples comparison.

\textnb{Performance Analysis:}
Fig.~\ref{res:online-boutique-dpu-efficiency} (1) shows RPS of three chains (`Home Query', `ViewCart', `Product Query'), each of which incur more than 11 data exchanges between functions.
NightCore's RPS remains 5.1\X$\sim$20.9\X lower than \name
even though it places all functions on the same node to fully exploit shared memory processing. 
We attribute this gap to the duplicated overhead of HTTP/TCP processing at both the worker node and cluster ingress. 
SPRIGHT, working with DPDK-based \textit{F-Ingress}, achieves up to 8.6\X RPS improvement vs. NightCore.
The same observation can be made by comparing \texttt{FUYAO-F} and \texttt{FUYAO-K}, which aligns with our findings in \S\ref{eval:cluster-ingress}.
Both, \name(CNE) and \name(DNE) take one step further: terminating TCP connections at the ingress and converting traffic to RDMA for fast transmission, which significantly reduce transport protocol processing overhead. Jointly working with the lock-free inter-node data transfers, using two-sided RDMA, \name (DNE) achieves 2.1\X$\sim$4.1\X improved RPS compared to \texttt{FUYAO-F} and 2.4\X$\sim$4.1\X improved RPS compared to SPRIGHT.

\name's DNE also outperforms \name(CNE) (1.3\X$\sim$1.8\X higher RPS) when handling more than 20 clients: since CNE co-locates user functions on the CPU, DOCA Comch is no longer needed. We change the CNE to  use eBPF's \skmsg to directly interact with the user functions for intra-node IPC. However, \skmsg's interrupt-driven model begins to experience interrupt processing load~\cite{receive-livelock} of CNE at high concurrency (also the single CNE is shared by \skmsg of many functions), throttling CNE's I/O performance. In contrast, DNE design offloads the I/O handling of user functions to the DPU, eliminating excessive interrupts and sustaining higher throughput. Under light load (fewer than 20 clients), CNE attains slightly lower latency than DNE, as interrupt overhead is minimal, and the CPU's general-purpose cores yield better responsiveness (see Table~\ref{tab:online-boutique-latency}). Both \name DNE and CNE latency have significantly lower than the other alternatives.

\begin{table}[b]
\vspace{-4mm}
\centering
\caption{Average latency (in ms) of online boutique chains.}
\vspace{-3mm}
\label{tab:online-boutique-latency}
\resizebox{\columnwidth}{!}{%
\begin{tabular}{l|lll|lll|lll}
\toprule
\textbf{}   & \multicolumn{3}{c|}{\textbf{Home Query}} & \multicolumn{3}{c|}{\textbf{View Cart}} & \multicolumn{3}{c}{\textbf{Product Query}} \\
\midrule\midrule
\textbf{\# clients}   & \textbf{20} & \textbf{60} & \textbf{80}  & \textbf{20} & \textbf{60} & \textbf{80}  & \textbf{20} & \textbf{60} & \textbf{80}  \\ \hline
\textbf{\name (DNE)}  &\textib{1.12}&\textib{2.55}&\textib{3.19} & 2.07        &\textib{3.35}&\textib{4.36} & 2.07        &\textib{3.92}&\textib{5.18} \\ \hline
\textbf{\name (CNE)}  & 1.43        & 4.39        & 5.62         &\textib{1.85}& 5.14        & 6.5          &\textib{2.06}& 6.04        & 7.63         \\ \hline
\textbf{FUYAO-F}      & 3.53        & 5.96        & 7.53         & 3.43        & 7.16        & 9.02         & 4.26        & 8.23        & 10.66        \\ \hline
\textbf{FUYAO-K}      & 4.88        & 8.28        & 20.5         & 4.95        & 9.22        & 10.18        & 5.33        & 9.46        & 11.39        \\ \hline
\textbf{SPRIGHT}      & 2.66        & 7.78        & 10.4         & 2.69        & 7.95        & 10.54        & 3.27        & 9.62        & 12.78        \\ \hline
\textbf{NightCore}    & 10.77       & 32.4        & 42.8         & 22.2        & 65.1        & 89.46        & 18.4        & 55.2        & 73.4         \\
\bottomrule
\end{tabular}
}
\end{table}

\subsubsection{Efficiency of DPU Offloading}
To assess the efficiency of DPU-offloaded network engine in \name, 
we analyze the CPU/DPU core utilization at the corresponding load level.
Other alternatives run their network engines on the CPU, similar to \name's CNE.

Fig.~\ref{res:online-boutique-dpu-efficiency} (4) - (6) shows the efficiency results.
FUYAO (both \texttt{FUYAO-K} and \texttt{FUYAO-F}) constantly saturates its assigned CPU core (more than 500\%) at high load. This is due to its reliance on one-sided write, which requires the receiver to continuously poll for data arrivals and takes up one core each on every worker node. In additional, FUYAO's reliance on kernel networking to terminate HTTP/TCP traffic from external clients and receiver-side copying
causes FUYAO's variants to have up to 3.5\X more CPU utilization than \name.

Looking at \name (DNE), its DNE operates in a run-to-completion busy loop, maintaining 100\% utilization of the assigned \textit{wimpy} DPU core regardless of the load (a total of 200\% since we use two worker nodes - thus two DNEs). Despite this, \name (DNE) consistently delivers superior performance than all the CPU-based alternatives.
This demonstrates that, through careful optimization (two-sided RDMA, run-to-completion loop and cross-processor shared memory), we can unleash the power of DPU even though 
its core is much less capable than the CPU core.
Although \name's DNE constantly utilizes the assigned DPU core at 100\%, the DPU core (Armv8 A72) runs at a lower clock speed (2.0 GHz) compared to the x86 CPU core (3.7 GHz), it compensates somewhat through a streamlined instruction set.
This highlights the effectiveness of leveraging a dedicated DPU for network processing, freeing up valuable CPU resources while maximizing overall system efficiency.

\vspace{-3mm}
\section{Conclusion}
\vspace{-1mm}
\name demonstrated the effectiveness of RDMA-capable DPU offloading in multi-tenant serverless clouds for improving data plane performance while reducing CPU usage.
By combining two-sided RDMA with local shared memory processing and early termination of HTTP/TCP protocol at the cluster ingress, \name achieves up to 20.9\X RPS improvement and 21\X latency reduction compared to state-of-the-arts serverless data planes, using either RDMA or shared memory, when serving a complex online boutique workload.

With \name, we show that these SmartNIC-like DPUs are Not in name only but have the ability to reshape the serverless data plane by delivering zero‐copy's efficiency and performance, while enforcing network isolation at no extra cost.
Through careful optimization of the CPU-DPU datapath (via CPU-DPU memory sharing) and the streamlined run-to-completion packet processing pipeline, a DPU-enabled network engine can largely outperform a CPU-based network engine (up to 1.8\X RPS improvement in an apples-to-apples comparison) when executing key data plane tasks in \name, including inter-node RDMA data transfers and managing multi-tenancy, while freeing up 7 CPU cores and actively using only two wimpy DPU cores.
\textnit{We will open-source the implementation of \name.}

\bibliographystyle{ACM-Reference-Format}
\bibliography{ref}

\end{document}